\documentclass{aastex}
\usepackage{emulateapj5}

\shorttitle{Metallicities and ages in Leo\,II}
\shortauthors{Koch et al.}

\begin{document}

\title{Complexity on Small Scales II: \\ Metallicities and Ages in the Leo II
Dwarf Spheroidal Galaxy}

\author{Andreas Koch\altaffilmark{2}, Eva K.~Grebel\altaffilmark{2}, 
        Jan T.~Kleyna\altaffilmark{3}, 
	Mark I.~Wilkinson\altaffilmark{4}, \\
	Daniel R.~Harbeck\altaffilmark{5},  
	Gerard F.~Gilmore\altaffilmark{4}, Rosemary F.G.~Wyse\altaffilmark{6}, 
	and N.~Wyn Evans\altaffilmark{4}}
\email{koch@astro.unibas.ch}

\altaffiltext{1}{Based on observations collected at the European Southern 
Observatory at Paranal, Chile; Large Programme proposal 171.B-0520(A), and 
on observations made through the Isaac Newton Groups' Wide Field Camera 
Survey Programme with the  Isaac Newton Telescope operated on the island 
of La Palma by the Isaac Newton Group in the Spanish Observatorio del 
Roque de los Muchachos of the Instituto de Astrofisica de Canarias.}
\altaffiltext{2}{Astronomical Institute of the University of Basel,
Department of Physics and Astronomy, Venusstrasse 7, CH-4102 Binningen, 
Switzerland}
\altaffiltext{3}{Institute for Astronomy, University of Hawaii, 2860 
Woodlawn Drive, Honolulu, HI 96822}
\altaffiltext{4}{Institute of Astronomy, Cambridge University, Madingley 
Road, Cambridge CB3 0HA, UK}
\altaffiltext{5}{University of Wisconsin, Department of Astronomy,
55343 Sterling, 475 N.\ Charter St., Madison, WI 53706-1582}
\altaffiltext{6}{The John Hopkins University, 3701 San Martin Drive, 
Baltimore, MD 21218}

\begin{abstract}
We present metallicities and ages for 52 red giants in the remote
Galactic dwarf spheroidal (dSph) galaxy  Leo\,II.  These stars
cover the entire surface area of Leo\,II and are radial velocity
members.  We obtained medium-resolution multi-fiber spectroscopy with
the FLAMES multi-object spectrograph as part of a Large Programme with
the Very Large Telescope at the European Southern Observatory, Chile.
The metallicities were determined based on the well-established
near-infrared \ion{Ca}{2} triplet technique. 
This allowed us to achieve a mean random error 
of 0.16\,dex on the metallicities, 
while other systematic effects, such as unknown variations in the dSph's 
[Ca/Fe]-ratio,  
may introduce a further source of uncertainty of the order of 0.1\,dex. 
The resulting metallicity
distribution is asymmetric and peaks at [Fe/H] $= -1.74$\,dex on the
Carretta \& Gratton scale.  The full range in metallicities extends
from $-2.4$ to $-1.08$\,dex.
 As in other dSph galaxies, no extremely
metal-poor red giants were found.  We compare Leo\,II's observed
metallicity distribution 
with model predictions for several other Galactic dSphs from the
literature.  Leo\,II clearly exhibits a lack of more metal poor stars,   
in analogy to the classical G-dwarf problem, which may indicate a comparable ``K-giant 
problem''. 
Moreover, its
evolution appears to have been affected by galactic winds.  We use our
inferred metallicities as an input parameter for isochrone fits to
Sloan Digital Sky Survey photometry of our target stars and derive
approximate ages.  The resulting age-metallicity distribution covers
the full age range from 2 to about 15 Gyr on our adopted isochrone scale. 
During the first $\sim 7$ 
Gyr relative to the oldest stars the metallicity of Leo\,II appears to have remained almost
constant, centering on the mean metallicity of this galaxy.  
The almost constant
metallicity at higher ages and a slight drop by about 0.3 dex thereafter 
may be indicative of rejuvenation by low metallicity gas. 
Overall, the age-metallicity relation appears to support the formation
of Leo\,II from pre-enriched gas.  
Evidence for enrichment is seen
during the recent 2 to 4 Gyr.  
Our findings support earlier derived photometric
findings of Leo\,II as a galaxy with a prominent old population and
dominant intermediate-age populations.   We do not see a significant
indication of a radial metallicity gradient nor age gradient in our
current data.  
\end{abstract}


\section{Introduction}

The majority of the galaxies in the Local Group are dwarf spheroidal
(dSph) galaxies, the least massive, least luminous type of galaxies
known.  They are mainly found within less than 300 kpc of massive
galaxies like the Milky Way or M31 (see, e.g., Grebel, Gallagher, \&
Harbeck 2003; van den Bergh 1999a).  The origin and nature of these seemingly
dark-matter-dominated dwarf galaxies is still poorly understood and is the 
subject of intense research.  DSphs are possibly the most
dark-matter-dominated objects in the Universe and may share a common
halo mass scale of $4 \cdot 10^7$ M$_{\odot}$ (e.g., Gallagher \&
Wyse 1994; Mateo et al. 1998; Odenkirchen et al.\ 2001; Kleyna et al.\ 2001; Klessen,
Grebel, \& Harbeck 2003; Wilkinson et al.\ 2004, 2006a).  This makes it
particularly interesting to understand their evolutionary histories.

The dSph census in the Local Group remains incomplete owing to the
intrinsically low surface brightnesses of these gas-deficient
galaxies.  The last three years alone saw the discovery of 13 new
Local Group dSph candidates (Zucker et al.\ 2004, Zucker et al.\ 2006a, 2006b, 
2006c; Willman et al.\ 2005; Belokurov et al.\ 2006a, 2006b, Martin et al. 
2006), seven of which
are new Milky Way companions.  This increases the number of known
Galactic dSph satellites to 15, omitting accreted companions and
additional objects whose nature remains yet to be confirmed (Grebel,
Gallagher, \& Harbeck 2006).  More than 75\% of {\em all} Milky Way
companions are found within 150 kpc; only three additional dSphs
believed to be Galactic satellites are found at larger distances.
Leo\,II, the dSph studied in the current paper, is one of the latter.  

Leo\,II was discovered by Harrington \& Wilson (1950).  A number of
mainly photometric studies were conducted in the subsequent decades.
Earlier studies based on resolved color-magnitude diagrams (CMDs)
placed Leo\,II at a distance of $\sim 230$ kpc (e.g., Demers \& Harris
1983; Bellazzini, Gennari, \& Ferraro 2005), $\sim 215$ kpc (Demers \&
Irwin 1993; Lee 1995), and $\sim 205$ kpc (Mighell \& Rich 1996).
These studies used different values for the (low) amount of Galactic
foreground reddening, different mean metallicities, and different
methods to derive the distance of Leo\,II, e.g., the apparent
luminosity of the horizontal branch or of the tip of the red giant
branch (TRGB).  In the current study, we adopt the TRGB distance of
$233 \pm 15$ kpc obtained by Bellazzini et al.\ (2005), which excludes
the contribution of luminous asymptotic giant branch (AGB) stars and
also shows consistency with the horizontal branch (HB) locus.  If this
distance is representative of Leo\,II's distance from the Milky Way
during most of its existence, it should have been little affected by
tidal stripping and other environmental effects exerted by the Galaxy.   

Leo\,II is a dSph galaxy with prominent old and intermediate-age
populations.  The presence of old populations was first established by
Swope (1967), who reported the detection of a red HB and of RR Lyrae
variables.  Deeper CCD exposures by Demers \& Irwin (1993) revealed
the existence of a less well populated blue HB.  The most
comprehensive variability study is that of Siegel \& Majewski (2000),
which presents 148 RR Lyrae stars and four anomalous Cepheids.  The RR
Lyrae stars indicate a dominant metallicity of $-1.9$ dex with a
metal-poor tail out to $-2.3$ dex.  The presence of intermediate-age
populations was first inferred from the detection of luminous carbon
stars in Leo\,II (Aaronson, Olszewski, \& Hodge 1983; Aaronson \&
Mould 1985, Azzopardi et al.\ 1985).  Five out of the seven C stars
and C star candidates probably belong to an intermediate-age
population judging from their position above the TRGB, although the
colors and luminosities of these stars suggest that they may be 3--4
Gyr older than the C stars detected in Leo\,I (see Lee 1995 for
details).  Moreover, a number of luminous AGB stars and a red clump
were detected in Leo\,II, which demonstrate the presence of an
intermediate-age population beyond any doubt (Lee 1995; Mighell \&
Rich 1996).  Hence the stellar age tracers show that Leo\,II
experienced an extended star formation (SF) history. 

The SF history of Leo\,II was quantified using deep Hubble Space
Telescope (HST) imaging extending below the old main-sequence turnoff
of this galaxy and theoretical isochrones.  Mighell \& Rich (1996) found a
mean age of $9\pm1$ Gyr with SF lasting for more than 7 Gyr.  They
conclude that about half of the stars in Leo\,II formed between 8 and
12 Gyr ago.  The oldest stars in Leo\,II are as old as the oldest
Galactic globular clusters, a common feature for all Galactic dSphs
and other nearby galaxies (Grebel \& Gallagher 2004).  Hernandez,
Gilmore, \& Valls-Gabaud (2000) suggest that the peak SF activity
occurred $\sim 8$ Gyr ago and that SF ended $\sim 6$ Gyr ago.  They
excised the HB populations from their CMD analysis, and their SF
history does not reproduce the old populations of Leo\,II.  Dolphin's
(2002) CMD-fitting algorithm reveals an extended SF history starting
some 15 Gyr ago and extending until 2 to 4 Gyr ago, although very
little activity occurred within that age bin.  The main SF activity
takes place in the 8 to 15 Gyr bin according to Dolphin (2002).  All
three photometric analyses agree in finding an extended SF history for
Leo\,II that lasted many Gyr, and that the main SF activity occurred
at relatively high ages of 8 Gyr or more.

In a recent,  shallower ground-based study Bellazzini et al.\ (2005)
discuss the detection of red giant branch bumps in Leo\,II.  The
luminosity of the main bump indicates that the majority of the stars
in Leo\,II is $\sim 4$ Gyr younger than regular Galactic globular
clusters of comparable low metallicity to the dominant stellar population 
in Leo\,II.
They also demonstrate that the red HB and red clump stars
in Leo\,II are more centrally concentrated than the ancient HB
population, akin to the population gradients found in many dSphs
(e.g., Stetson, Hesser, \& Smecker-Hane 1998; Hurley-Keller, Mateo, \&
Grebel 1999; Da Costa et al.\ 2000; Harbeck et al.\ 2001).    

Interestingly, in all three deep HST CMD studies the isochrone fits
require a relatively high metallicity of $\sim -1.6$ dex or even $\sim
-1.1$ dex.  Indeed, when comparing isochrones for old populations to
globular clusters with metallicity determinations on the traditionally
used Zinn \& West (1984) scale, the isochrones favor too high a
metallicity for metal-poor populations (Grebel 1997, 1999).  As
discussed by Dolphin (2002), this can be in part circumvented by using
the Carretta \& Gratton (1997) scale for globular clusters instead
(see also discussion in Bellazzini et al.\ 2005).  Nevertheless, the
age-metallicity degeneracy that plagues purely photometric derivations
of SF histories from resolved stellar populations remains.       

The age-metallicity degeneracy can be resolved through spectroscopic
metallicities.  The first spectroscopic metallicity determination for
Leo\,II was carried out by Suntzeff et al.\ (1986), who measured three
red giants in this galaxy and found [Fe/H] $\sim -1.9$ dex.  Their
metallicity scale corresponds to the Zinn \& West (1984) scale.  More
recently, Bosler et al.\ (2004) spectrometered 41 red giants in Leo\,II
and found a mean metallicity of $-1.57$ dex with a spread ranging from
$-1.26 >$ [Fe/H] $> -2.32$ on the Carretta \& Gratton (1997) scale.

In our current paper, we present spectroscopic metallicity
measurements for 52 red giants.  \textsection 2 introduces the data
and briefly describes the reduction strategy. In \textsection 3 the
calibrations of the measured CaT equivalent widths (EWs) onto a
reference metallicity scale are presented and compared to existing
data sets.  The resulting metallicity distribution function (MDF) is
analyzed in \textsection 4 and \textsection 5 discusses this MDF in
the context of simple models of chemical evolution. In \textsection 6
we turn to the derivation of stellar ages and to the age-metallicity
relation of Leo\,II.  Finally, in \textsection 7 we summarize our
findings.

\section{Observations and reduction}

Our observations of Leo\,II were taken as part of the European Southern Observatory (ESO) Large
Programme 171.B-0520(A) (principal investigator: G.F.\ Gilmore; see
Koch et al.\ 2006a, 2006b; Wyse et al.\ 2006; Wilkinson et al.\ 2006b
for more information), which aims at elucidating the kinematic and
chemical characteristics of Galactic dSphs. 
Five fields in Leo\,II were observed with the FLAMES spectrograph
(Pasquini et al.\ 2002) at ESO's Very Large Telescope (VLT).  FLAMES
was used with the GIRAFFE multi-object spectrograph in low-resolution
mode ($R\sim6500$).  The GIRAFFE spectrograph was fed by the MEDUSA
fiber system, which provides up to 132 fibers per exposure.  In the
same Large Programme, we also analyzed the metallicities of the Carina
dSph, which is described in detail in Koch et al.\ (2006a, hereafter
Paper I).  We refer the reader to Paper\,I for details on observing
strategy, data reduction, and calibration techniques, which will be
briefly summarized in the following.  Both for Carina and for Leo\,II
the L8 grating was used, centered on the near-infrared Ca\,II triplet
(CaT) and covering a wavelength range of 820.6 nm to 940 nm.

\subsection{Target selection and acquisition}

The wide field of view of the FLAMES instrument has a diameter of
25$\arcmin$.  In principle, this enables one to cover stars in the
entire projected area of the Leo\,II dSph with one single pointing,
since its nominal tidal radius is 8$\farcm$7 (Irwin \& Hatzidimitriou
1995).  However, in order to account for crowding, varying stellar
density, and fiber collisions, and also in order to optimize the
sampling out to large radii near and beyond the tidal radius, we
observed five different (overlapping) fields in several fiber 
configurations (see Tables 1 and 2).

Our target stars were selected using photometry and astrometry
obtained by the Cambridge Astronomical Survey Unit (CASU; 
Irwin \& Lewis 2001; McMahon et al. 2001) at the 2.5~m
Isaac Newton Telescope (INT) on La Palma, Spain.  From these data we
selected red giant candidates covering the luminosity range from the TRGB region
($V \sim 18.5$) down to 2.5~mag below the TRGB ($V \sim 21$).  As the
CMD in Fig.~1 shows we furthermore aimed at covering the full width of
the red giant branch (RGB) to ensure that also potential extremely
metal-poor and metal-rich red giants were included and in order to
minimize any bias with regard to age or metallicity.  In total, we
selected and observed 195 red giant candidates. 

For the purpose of accurate metallicity determinations based on the
CaT, we targeted four Galactic globular clusters, which will
serve as a reference scale for our measurements later on (Rutledge et
al.\ 1997b; Paper\,I). Red giants in these clusters were selected from
the standard fields of Stetson (2000), which provide $B$ and $V$-band
magnitudes in the Johnson-Cousins system as defined by Landolt's
(1992) $UBVRI$ standard stars.  In contrast, the CASU $g',\,r',\,i'$-photometry was
obtained using filters mimicking the SDSS $g\,r\,i$ system (Fukugita
et al.\ 1996).  Since our targeted region on the sky is partially
covered by a standard star field in Leo\,II (Stetson 2000), we can
directly compare both sets of photometry and place both the
calibration clusters and the Leo\,II targets for the FLAMES run on
Stetson's homogeneous photometric standard system.  This avoids any
photometric offset, which may ultimately result in a incorrect calibration
of the CaT scale as this scale is defined for $V$-band photometry
(Armandroff \& Da Costa 1991).  Via a linear transformation we
obtained 
\begin{eqnarray}
V_{JC}\,=\,g'\,-\,0.549\times(g'-r')\,-\,&0.727& \\
(\mathrm{18\,\,stars,\,\,}\sigma = &0.025)&.  \nonumber 
\end{eqnarray}

For the purpose of age determinations once metallicities have been
measured, it is desirable to obtain as accurate a set of photometry
as possible in a suitable set of filters. Since the INT filter system 
used in the CASU\footnote{see \url{http://www.ast.cam.ac.uk/$\sim$wfcsur/}.} 
parallels the SDSS system (Karaali et al.\ 2005; Jordi, Grebel, \&
Ammon 2006) and since Leo\,II is also covered by the fifth data release (DR5) of the 
Sloan Digital Sky
Survey\footnote{
The data from the SDSS DR5 are publically available under \url{http://www.sdss.org/dr5/}.} 
(SDSS; e.g., Stoughton et al.\ 2002; Adelman-McCarthy et al.\
2006), we matched the CASU photometry of our targets with the (less
deep) SDSS photometry.  SDSS photometry
is obtained by driftscan techniques, is very well calibrated, and
offers a highly homogeneous data set.  Moreover, the theoretical
isochrones of the Padova group have been transformed into the SDSS
photometric system (Girardi et al.\ 2004), facilitating age estimates 
via isochrone fitting.  We determined the following transformation 
relations for the conversion of CASU to SDSS photometry: 
\begin{eqnarray}
g\,=\,g'\,+\,0.127\times(g'-r')\,-\,&0.814& \\	
i\,=\,i'\,+\,0.114\times(r'-i')\,-\,&0.385& \\
(\mathrm{1700\,\,stars,\,\,} \sigma = &0.15)&\nonumber.
\end{eqnarray}
Here the standard SDSS photometric system is denoted by  $g$ and $i$, 
as is the usual notation, while the CASU magnitudes are given by the primed 
$g'$, $r'$ and $i'$.  

As mentioned already, the spectroscopic observations were carried out
using the same strategy and instrumental setups as described in
Paper\,I.  The location of our fields and the dates of observation are
listed in Tables 1 and 2.  Although we aimed at exposing each
configuration for 6\,hrs to reach nominal signal-to-noise (S/N) ratios
of at least 20, which would enable us to derive highly accurate EWs at
our spectral resolution of $\sim 6500$, a large fraction of our nights
was hampered by sky conditions with a seeing as bad as 2$\arcsec$.  
Therefore, the
median S/N actually achieved after processing the spectra was only  
15\,pixel$^{-1}$.  Altogether 197 targets were observed, whose
individual positions are shown in Fig.~2.

\subsection{Data reduction}

Details of the reduction process are given in Paper\,I. In essence, we
used version 1.09 of the FLAMES data reduction system, girbldrs, and
the associated pipeline version 1.05 (Blecha et al.\ 2000).  After
standard bias correction and flatfielding, the spectra were extracted
by summing the pixels along a slit of 1\,pixel width.  The final
rebinning to the linear wavelength regime was done using Th-Ar
calibration spectra taken during daytime. 

Sky subtraction was facilitated by the allocation of about 20 fibers 
configuration to blank sky. 
By subtracting the average sky spectrum from the science
exposures using IRAF's{\footnote{IRAF is distributed by the National
Optical Astronomy Observatories, which are operated by the Association
of Universities for Research in Astronomy, Inc., under cooperative
agreement with the National Science Foundation.}} {\em skytweak}, we
obtained an accuracy of the final sky-subtracted spectra of the order
of 2\%, taken as the 1$\sigma$-dispersion of the medians of the
sky-subtracted spectra in any exposure divided by the median sky. 

Our data set was then completed by co-adding the dispersion-corrected
and sky-subtracted science frames, weighted by the exposures'
individual S/N, and subsequent rectification of the continuum.  Fig.~3
shows representative spectra across our magnitude range for which
reliable EWs could be determined.

\subsection{Membership estimates}

In order to separate Leo\,II's RGB stars from Galactic foreground
stars, we determined the individual radial velocities of each target
star by means of cross-correlation of the three calcium triplet lines against synthetic
template spectra using IRAF's FXCOR task. The template was synthesized adopting 
representative equivalent widths of the CaT in red giants. The typical 
median velocity error achieved in this way lies at 2.4\,km\,s$^{-1}$.
This method will be laid out in detail in a
forthcoming paper focusing on the dynamical aspects of Leo\,II
(Kleyna et al.\ 2006, in prep.).  

At a systemic velocity of (76\,$\pm$\,1.3)\,km\,s$^{-1}$ (Vogt et al.\
1995), Leo\,II's velocity distribution will inevitably contain 
a number of Galactic foreground stars along
the galaxy's line of sight. However,
owing to our selection of target stars whose colors and luminosities
are consistent with membership in Leo\,II and due to the much lower
Galactic field star density, Leo\,II's velocity peak clearly stands
out against the Galactic contribution (see the histogram in Fig.~4). 
 By application of the Besan\c con synthetic Galaxy model 
(Robin et all. 2003) we find a Galactic distribution with a mean and dispersion 
of 15\,km\,s$^{-1}$ and 57\,km\,s$^{-1}$. From this distribution 
we estimate that the number of Galactic interlopers, subject to 
our color-magnitude criteria, within 3\,(5)\,$\sigma$ of 
an initial fit of a Gaussian velocity peak to the Leo\,II data with mean and dispersion of 
79.8\,km\,s$^{-1}$ and 
7.3\,km\,s$^{-1}$, 
amounts to 2\,(4) and thus should not give rise to any concern in our analysis 
of Leo\,II's stellar populations.

Rejecting 23 apparent 
radial velocity non-members that deviate more than 5\,$\sigma$ from 
this initial fit, we approach
the mean heliocentric velocity and line-of-sight velocity dispersion
by performing an iterative error-weighted maximum-likelihood fit assuming a
Gaussian velocity distribution.  

The resulting mean of (79.2$\,\pm\,$0.6)\,km\,s$^{-1}$ and velocity
dispersion of (6.8$\,\pm\,$0.7)\,km\,s$^{-1}$ are in very good
agreement with Vogt et al.\ (1995).  These authors found a a central
velocity dispersion of (6.7$\,\pm\,$1.1)\,km\,s$^{-1}$ based on 31
high-resolution spectra of stars within the core radius.  Depending on
the criteria for defining membership we end up with 166 radial
velocity members if we adopt a $3\sigma$-cut in the velocity
distribution, whereas a more conservative cut at $\pm 2\sigma$ leaves
155 apparent member stars. Given the low likelihood of interlopers
in our sample, we will apply  the $3\sigma$-cut in the following.

\section{Metallicity calibration}

The infrared lines of the singly ionized calcium ion at 8498, 8542,
and 8662\,\AA~are among the strongest absorption features in the
spectra of RGB stars.  Their EWs can thus be accurately determined
and reach values of typically a few \AA ngstr\"oms. As was nicely
demonstrated in the pioneering works of Armandroff \& Zinn (1988);
Armandroff \& Da Costa (1991) and Rutledge et al.\ (1997a,b), the CaT
can be employed to measure the  metallicity of red giants in old and
metal-poor populations.  
Subsequently, the CaT method was extended to and calibrated for the
entire range of metallicities of $-2 <$ [Fe/H] $< -0.2$ and for ages
covering $2.5 <$ age [Gyr] $< 13$ by Cole et al.\ (2004).  This
fundamental work permits us to apply the CaT method to populations
that also contain stars of intermediate ages, such as Carina or
Leo\,II. 

The EWs of the CaT lines were measured using a modified version of
Da Costa's EWPROG code kindly made available by A.~A. Cole (see also
Cole et al.\ 2004).  Each of the lines was fit by a Gaussian plus a
Lorentzian component in the line bandpasses (and associated continuum)
as defined in Armandroff \& Zinn (1988).  The final EW was then
obtained by summing up the flux in the theoretical profile across the
bandpass.  The corresponding uncertainties were determined from the
residuals of the fit.

In order to obtain a self-consistent analysis of the Leo\,II data, we
rederived the relation between line strengths and metallicity. We
follow the original prescription of Rutledge et al.\ (1997a, hereafter
R97a) in defining the Ca line strength as the weighted sum of the three
CaT lines:
\begin{equation} 
\Sigma W = 0.5 \, W_{8498} + W_{8542} + 0.6 \, W_{8662} 
\end{equation}
This definition differs from that employed throughout Paper\,I, where 
we made no use of the first of the Ca lines ($W_{8498}$). 
Owing to the generally low S/N of our Leo\,II spectra, we 
were able to measure EWs of all the three CaT lines for only 20 radial 
velocity members (down to a S/N of approximately 10), 
while we could reliably fit the two stronger lines at
$\lambda\lambda\,$8542, 8662 in a further 32 stars (down to S/N $\sim$ 6). 
  In order to place the
linestrengths derived from these spectra on the same scale as the ones
with better data quality, we established a linear relationship between
the linestrength $\Sigma W$ and the strength using only the two
stronger lines from the 72 high S/N spectra (with S/N ratios of about 60--80) 
of our calibration clusters, NGC\,3201, NGC\,4147, NGC\,4590 and NGC\,5904 
(for details on these  
calibration clusters, we refer to Paper\,I.)  We find the linear 
relation
\begin{equation} 
\Sigma W = 1.13 \, (W_{8542} + 0.6 \, W_{8662}) + 0.04, 
\end{equation}
with associated uncertainties of 0.02 on each of its coefficients 
and a r.m.s. scatter of 0.05\AA.

The power of the CaT calibration lies in the introduction of a {\em
reduced} width, $W'$, which reduces the strong effects of stellar
gravity on the Ca line strengths to first order (and the far lesser
influence of effective temperature). This relation applies within 
a given cluster, for which V$_{HB}$ is the apparent magnitude of the 
horizontal branch.
The reduced width is defined in the
high-S/N work of R97a as 
\begin{equation} 
W' = \Sigma W + \beta\,(V-V_{HB}), 
\end{equation}
where $V$ denotes the stellar magnitude. Based on our observed high
S/N spectra of the four calibration clusters, we find the slope in this
relation to be $\beta=-0.55\,\pm0.08$\AA\,mag$^{-1}$, which is
somewhat shallower than the canonical value of $-0.64\pm 0.02$.  
The quoted error on $\beta$ is obtained from the 
formal uncertainty from the fit, multiplied with the median error 
of unit weight (m.e.1), which is defined by summing over all 
stars in the four clusters as 
m.e.1\,=\,$\left( \sum \varepsilon^2/\sigma^2 \right)^{0.5}  \, \nu^{\,-0.5}$. 
Here $\varepsilon$ denotes the deviation of each star's $\Sigma W$ 
from the fit, $\sigma$ is the measured uncertainty in the EWs, and $\nu$ 
is the number of stars in the fit, minus the number of calibration clusters plus one 
(see also R97a). From our data we find a m.e.1 of 2.56. 

For the
HB level $V_{HB}$ of Leo\,II, we adopted the value of 
22.17$\pm$0.14\,mag (Siegel \& Majewski 2000), which refers to the
HB locus based on the analysis of a sample of RR\,Lyrae stars. 

\subsection{Uncertainties affecting the metallicity measurements}

There are various metallicity scales to be found in the literature
(e.g., ZW, CG, Kraft \& Ivans 2003).  The CG scale and the Kraft \&
Ivans scale are  based on high resolution data, but there is no
physical reason to give preference to any of them.  Traditionally the
ZW has been the most widely used scale, but more recent calibrations
of the CaT method have mainly used the reference metallicity scale of
CG as introduced by Rutledge et al. (1997b).   In our current study, we adopt this scale
as well.  The final calibration of the reduced width onto metallicity
is then obtained via the linear relation 
\begin{equation} 
[Fe/H]_{CG} = (-2.85 \pm0.09) + (0.43\pm0.03)\,W',  
\end{equation}
with an r.m.s scatter of 0.02\,dex (and m.e.1\,=\,0.67), as established from our four calibration clusters. 
We note that the coefficients in eq.~7 differ slightly from the calibration 
established in Paper\,I, which can be attributed to different 
definitions of the CaT linestrength $\Sigma W$. Nevertheless, reassuringly both calibrations   
 agree to within their uncertainties. 
Fig.~5 shows the distribution of our target stars in the $W$ vs. $V-V_{HB}$ plane, 
together with some isometallicity lines according to eqs. 6,7 to guide the eye. 
This diagram already indicates that there is a wide range of metallicities present in Leo\,II, 
where the majority of stars lies at a [Fe/H] of about $-1.7$\,dex.  

The mean random error introduced by measurement errors in the EWs and our photometric uncertainties  
amounts to 0.06\,dex, while the full mean error on our metallicities, accounting for all 
calibrations (eqs.~5--7), is found to be 0.16\,dex. 

However, there are numerous additional potential
sources of uncertainty, which may affect the accuracy of our
calibration of Ca onto iron.  Among these are the [Ca/Fe] ratios for the stars in
dSphs, which, for the sake of the method, have to be assumed to be
comparable to the abundance ratios in Galactic globular clusters.  On the other hand,
high-resolution spectroscopic data of red giants in dSphs and dwarf
irregulars found generally lower [$\alpha$/Fe] ratios than in Galactic
populations of the same [Fe/H] (e.g., Hill et al.\ 2000; Shetrone,
C{\^o}t{\'e}, \& Sargent 2001; Fulbright 2002; Shetrone et al. 2003; Tolstoy et al. 2003; 
Geisler et al.\ 2005;
Pritzl et al.\ 2005).  However, our own high-dispersion analysis of
stars in the Carina dSph galaxy (Koch 2006) and the
comparison with CaT measurements of the same stars from Paper\,I shows
generally good agreement in the derived metallicities albeit with some scatter. 

In this context it is worth noting that the recent high-dispersion
analyses by 
Bosler et al. (2006; hereafter B06) 
have allowed these authors to establish a calibration of CaT line
strength directly onto [Ca/H], yielding a median [Ca/H] ratio of
$-1.65$\,dex with stars covering the regime of $\sim-2.6$ to
$\sim-0.6$\,dex.  Under the simplifying assumption that all these
stars share an average metal content of [Fe/H]$\sim-1.9$\,dex, B06
tentatively estimate Leo\,II's global [Ca/Fe] ratio to be of the order of
$\sim+0.3$\,dex, similar to the old Galactic globular clusters and field stars.  
This apparently indicates that the 
calibration of the CaT onto metallicity, which assumes a [Ca/Fe] as in Galactic globular 
clusters, can safely be applied to the Leo\,II dSph without introducing any significant bias 
due to a strongly deviant Ca abundance ratio.

Other potential error sources also include the {\it a priori} unknown
variations of the HB level with both age and metallicity, which may
introduce uncertainties of the order of $\pm 0.05$ dex (Da Costa \& Hatzidimitriou 1998; 
Cole et al.\ 2004), which is below the precision of the abundance measurements themselves.  
These effects are discussed in detail in Paper\,I. 
 
Hence, in the light of
the comparatively low S/N of our spectra, we conclude that the 
measurement errors on the CaT EWs are the major contributors to the
quoted uncertainty, rather than the systematic effects such as 
variations in the HB or the [Ca/Fe]-ratios mentioned above, so that the former 
will effectively lead to a broadening of the
derived MDF.

\subsection{Comparison with other spectroscopic measurements for
individual stars}

Of the 74 red giants from the study of B06, which were observed with
the Low-Resolution Spectrograph (LRIS) at the Keck 10-meter telescope,
32 coincide with our targets.  Fifteen of these had sufficient S/N in
our data to allow us to determine metallicities\footnote{Note that
both spectrographs achieve comparable resolutions (FWHM) -- 1.55\AA\ in the case
of LRIS and 1.31\AA\ for our GIRAFFE data.}.  These common CaT measurements can be
employed for a cross check of the accuracy of the method and to
validate the respective measurements.  Fig.~6 shows the comparison of
the linestrengths $\Sigma$W (left panel) and the reduced widths (right panel) 
from both data sets.  A linear error-weighted
least-squares fit to the samples yields the relations
\begin{eqnarray} 
\Sigma W_{\mathrm{This\,work}} &=& (1.04\pm0.04)\,\Sigma W_{\mathrm{B06}} - (0.08\pm0.11)\\  
W'_{\mathrm{This\,work}} &=& (1.04\pm0.12)\,W'_{\mathrm{B06}} + (0.21\pm0.30).  
\end{eqnarray}
The fact that these are not one-to-one relations is to be expected,
since B06 measured EWs of the CaT lines by fitting a single Gaussian, whereas we used an 
additional Lorentzian to yield a better fit in the line wings (see also Cole et al. 2004).
Moreover, B06 
relied on the definition of $W'$ in terms of a slope $\beta$ 
from the literature (R97a),
whereas we obtained different estimates of $W'$ by using our 
own internal calibration, which finally 
reflects in the differences seen in either $\Sigma$W and W'. 
We note, however, that, with a median value of 23, B06 achieved S/N ratios slightly 
higher than ours.  
Generally, our 
analysis yields reduced widths that are larger by $\sim 0.2$\,\AA. 

\section{Leo\,II's metallicity distribution function and chemical 
evolution models}

\subsection{The metallicity distribution function}

Fig.~7 shows our final metallicity distribution (MDF) for 52 Leo\,II
red giants on the scale of CG.  This MDF is peaked at an
error-weighted mean metallicity of $(-1.74\pm0.02)$\,dex and exhibits
a full range in [Fe/H] of $-$2.4 to $-1.08$\,dex.  Note the
apparent lack of more metal-poor stars. 
The observed dispersion of the MDF amounts to 
0.29\,dex, but accounting for the measurement uncertainties by
subtracting these in quadrature, Leo\,II's internal metallicity
dispersion is estimated to be 0.23\,dex.  These values are in reasonable
agreement with the estimates of Bosler et al.\ (2004), who derived a
mean metallicity of  $-$1.57\,dex on the CG scale based on CaT spectra
of 41 red giants. 

All in all, our spectroscopic metallicities span about 1.3 dex in
[Fe/H].   As Fig.~8 implies, there is no apparent trend of metallicity and location in
the CMD seen in our data: both metal-rich and metal-poor giants are
found at the same location on the RGB. This indicates that there is
a comparably broad range of ages present in Leo\,II. In fact, Bosler
et al. (2004) have demonstrated that Leo\,II's RGB stars may span ages
between 2 and 13\,Gyrs.  We will address the question of an
age-metallicity relation in our data in Section 5.  The range of
metallicities and the mean metallicity measured by us are also in
excellent agreement with fiducial-based photometric metallicity
estimates on the CG scale by Bellazzini et al.\ (2005).  These authors
find a mean metallicity of $-1.74$ with a dispersion of 0.3 dex.  The
full range of metallicities that they derive from globular cluster
fiducials is $-2.4 <$ [Fe/H] $< -1.2$, also on the CG scale.  This
also agrees well with the estimates from Mighell \& Rich's (1996) HST
photometry (mean metallicity $-1.6$).  
 
The overall shape of the MDF is highly asymmetric:  
At
the metal-rich end it shows a smooth, but rapid, fall-off. A sharp
drop-off of an MDF towards higher metallicities is generally
attributed to a sudden cessation of SF, possibly caused by a strong
mass loss (e.g., through winds). Although we see a metal-rich fall-off
in our MDF, it is not as distinct in our data as, e.g., in the MDF of
Bosler et al. (2004).
Towards the metal-poor end, the MDF shows a noticeable number of metal-poor stars, but   it 
exhibits a sharp drop below $-$2\,dex. While we find seven stars with $-2.5\la$[Fe/H]$\la-2$, 
simple models of chemical evolution (e.g., Pagel 1997, p. 218) generally predict an 
excess of stars in this regime and at lower metallicities (see 
the model curve for G-dwarfs in Fig.~7). 
This is in analogy to the ``G-dwarf problem'' in the 
solar neighbourhood (e.g., Nordstr\"om et al. 2004). 
Since these quantitative models are based on the prescriptions for long-lived stars, 
they cannot be immediately applied to our data that are based on K-giants of finite, negligible 
lifetimes. However, under the assumption of a standard initial mass function, zero initial metallicity, 
and accounting for the loss of metals from the galaxy, in the simple model of chemical 
evolution one would qualitatively also expect an excess of the oldest and most metal-poor (K-) giants,  
so that we are here in fact faced with a ``K-giant''-problem. 
Assuming a constant SF rate, the oldest population should then contribute the largest fraction 
of K giants. 

\subsection{Comparison with 
chemical evolution
models for other dSphs}

We now turn to the question whether more information can be extracted
from the observed MDF when comparing it to sophisticated models,
which preferably incorporate the effects of outflows and infalling
gas.  A number of such models were calculated for dSph galaxies (e.g.,
Ikuta \& Arimoto 2002; Lanfranchi \& Matteucci 2003, 2004; Hensler et
al.\ 2004; Font et al.\ 2006).  
Carigi, Hernandez, \& Gilmore (2002) are the only group, who 
calculated chemical evolution models of Leo\,II, based on the SF
history derived by Hernandez et al.\ (2000). 
However, these models do not account for
the old population in this galaxy (see also Dolphin 2002).  Furthermore, Carigi et
al.\ (2002) do not provide model predictions for Leo\,II's stellar MDF
in the literature, yet for our current work, we need theoretical MDFs
to compare to. 

For convenience, we therefore chose to overplot the models by
Lanfranchi \& Matteucci (2004, hereafter LM04) on our data. Although
these authors did {\em not} model the evolution of Leo\,II itself, we
can derive some basic properties of this galaxy by comparing our
observed MDF  with the published theoretical MDFs for other dSphs. The 
modelled MDFs comprise the dSphs Carina, Draco, Sagittarius,
Sextans, Sculptor and Ursa Minor (Fig.~9). Each of the dSphs is
characterized by its SF rate and a wind efficiency,
the latter being the proportionality constant between gas mass and
SF rate.  These parameters were optimized by LM04 to fit extant 
observational data such as the [$\alpha$/Fe] ratios from red giants, their total mass, and
total gas mass. 
As discussed above, it is not a priori self evident that the model prescriptions based 
on long-lived stars can be immediately applied to observed K-giant MDFs. 
However, a comparison of the K-giant sample of the solar neighbourhood (McWilliam 1990) 
to the local G-dwarf MDF from Nordstr\"om et al. (2004) does not show   
any evidence for a significant difference, as both of them 
appear as a narrow distribution, peaked slightly below solar (see also Cole et al. 2005). 
Thus we continue by following  
the long-standing practice of comparing G-dwarf models and MDFs of K-giant as unbiased tracers of
low-mass stars.     

As Fig.~9 implies, all the shown MDFs are rather similar, with the exception of
Sgr. This is particularly interesting considering the wide range of absolute magnitudes 
covered by these galaxies (Mateo 1998). In this context, the mean abundances 
are comparatively similar and individual differences are basically reflected in 
the distributions' details, as we will discuss below. 
In terms of the location of its peak metallicity
Leo\,II's observed MDF resembles the model MDFs of the dSphs Car, Scl,
and UMi.  However, the probabilities for the same underlying
population derived from a Kolmogorov-Smirnov (K-S) test are 13\% for
UMi and Leo\,II, 29\% for Car and Leo\,II, and zero for all other
dwarfs shown in Fig.\ 9.  In this simplified sense, Leo\,II and Car
are the two most similar dSphs of the entire set. 

In contrast to Leo\,II, which shows an extended SF history from early times on 
that seems to have peaked around 9\,Gyr (e.g., Mighell \& Rich 1996; Dolphin 2002; 
Bosler et al.\ 2004), 
the dSph UMi is essentially a purely old galaxy (Carrera
et al.\ 2002; Wyse et al.\ 2002) and hence not a good comparison
object for Leo\,II, even though the ancient SF lasted several Gyr (see
Ikuta \& Arimoto 2002; Grebel \& Gallagher 2004).  The Car dSph, on
the other hand, has experienced ancient and intermediate-age SF, but
in contrast to Leo\,II this activity occurred in well-separated
episodes (e.g., Smecker-Hane et al.\ 1994 and references in Paper I).
In fact, it is the only dSph known to exhibit such clearly episodic
SF.  The bulk of the stars in Car seems to have formed some 7 Gyr ago
(Hurley-Keller et al.\ 1998).  In terms of the age range covered by
the SF in Car and Leo\,II, these two dSphs are similar.  However, they
show considerable differences in the details of their evolutionary
histories (see, e.g., Grebel 1999, 2000; van den Bergh 1999b, 2000).      

The overall low mean metallicities of dSphs (Grebel et al.\ 2003,
Table 1) are most easily explained by a low SF rate according to the
models of LM04, which would then leave a reservoir of gas at the final
epoch of SF. The fact that there is an observed termination of the SF
is then indicative of an efficient gas removal, e.g., via stripping by
the Galaxy (Ikuta \& Arimoto 2002).  It nonetheless remains difficult
to explain the {\em present-day} gas deficiency in dSphs (e.g.,
Gallagher et al.\ 2003), which cannot be caused by pure ram pressure
stripping in a smooth intergalactic medium (Grebel et al.\ 2003).

In order to gain insight in the processes governing the shape of the
MDF and the related evolutionary parameters, we shifted all LM04
models to the same peak metallicity as measured in Leo\,II (Fig.~10).
Thus one assumes the identical effective nucleosynthetic yield and similar SF  
efficiencies for all the
modeled systems.  The majority of the models, except for the
predictions for Sgr and UMi, tend to produce a steep decline of the
MDFs towards the metal-rich tail. This is caused by an intense wind,
which effectively drives out the metals and thus prevents any further
substantial enrichment or subsequent SF.  For the cases of Scl, Sgr,
and UMi, the drop in the MDF is not that sudden and approximates the
observations of Leo\,II fairly well.  The respective galactic wind
efficiencies are 13 (times the SF rate) for Scl, 9 for Sgr, and
10 for UMi.  The LM04 models tend to overestimate the number of
metal-poor stars, leading to the already mentioned G-dwarf problem (Sect.\
4.1).  This is most pronounced in the comparison of the models for Car
and Dra with Leo\,II's MDF (Fig.~10), while the Scl and Sgr model MDFs
resemble Leo\,II's metal-poor tail more closely. The latter coincidence 
is remarkable, as LM04 did not include any pre-enrichment in their models. 
In the context of the intense galactic winds it is worth noticing that 
the gas masses of these galaxies predicted  by the LM04-models (of the order of
3$\times$10$^{-4}\,$M$_{\rm tot}$) are lower than 
the upper limits placed by observations (Mateo 1998; Grebel et al. 2003), but still 
well consistent with the overall gas deficiency in dSph galaxies.

Overall, as one might expect none of the dSphs models gives a
perfect representation of Leo\,II's MDF since LM04's models were not
made to fit its MDF.  The K-S probabilities are below the 5\% level
with the exception of Scl (45\%), UMi (34\%), and Sgr(25\%).   Scl is
a predominantly old galaxy with subpopulations distinct in metallicity
and possibly in age and kinematics (Grebel, Roberts, \& van de Rydt
1994; Grebel 1996; Hurley-Keller et al.\ 1999; Majewski et al.\ 1999;
Harbeck et al.\ 2001; Tolstoy et al.\ 2004).  Likewise, UMi is an
ancient galaxy, but in spite of its extended  early SF it does not seem
to show metallicity gradients (e.g., Carrera et al.\ 2002).  
Sgr
formed stars over a very long period of time.  Intermediate-age
populations dominate (mean age of 8$\pm$1.5\,Gyr; Bellazzini et al.\
2006), but low-level SF may still have occurred until 0.5 to 1 Gyr ago
(Bellazzini, Ferraro, \& Buonanno 1999).  In fact, SF may have
occurred with a variable rate leading to multiple peaks (Layden \&
Sarajedini 2000).  The massive Sgr is an exceptional case among the
dSphs considered here since its evolution has been strongly affected
by its continuing disruption by the Milky Way (Ibata et al.\ 1994;
Majewski et al.\ 2003). 
No two dSphs share the same SF history or enrichment history, and in
addition external effects such as ram pressure or tidal stripping or
the early Galactic UV radiation field may have affected their
evolution (van den Bergh 1994; Grebel 1997).  Hence it may not be
surprising that Leo\,II's MDF is not well matched by models calculated
for other dSphs.  If Leo\,II's distance of $\sim$230\,kpc is
representative of the distance that it had during most of its
existence, then external effects may have played less of a role than
for the closer dSphs.  For instance, Siegel \& Majewski (2005) report
on only a marginal stellar component beyond the King (1962) tidal radius, 
which cannot be distinguished 
from statistical fluctuations.  
It is clear that
the detailed understanding of Leo\,II's MDF and evolution
will require detailed modeling adapted to the special properties of
this galaxy. However, based on the comparison with the LM04
models, we may speculate that also Leo\,II has been affected by 
winds, albeit at a modest level.

\section{Stellar Ages}

\subsection{Obtaining age estimates}

In order to gain a deeper insight into the detailed SF history of
Leo\,II we need to couple our spectroscopic abundances with age
estimates.  We can derive ages for all stars with metallicity
estimates through isochrone fitting.  For each of these stars we fit
its location in the CMD using  ($g-i$, $g$)-isochrones of  that star's individual
metallicity.  For this purpose, we employed a set of Padova isochrones
(Girardi et al.\ 2004) in the SDSS photometric system.  These were
interpolated to yield an estimate of each red giant's age.  

A few of our targets lie outside of the applicable isochrone grid. 
On the one hand this may be due to problems arising from the color transformations 
of theoretical isochrone models to the observational plane and 
the resulting failure to simultaneously and perfectly reproduce 
all features of a stellar system and in particular the RGB (e.g., 
Vandenberg et al. 2000;  Gallart, Zoccali \& Aparicio 2005). 
In the case of our data, this happens when our targets either turn out to 
be brighter than the 
respective theoretical tip of the RGB, or when they lie redwards of
the oldest available isochrone (cf. Tolstoy et al. 2003). 
The latter may be caused by
statistical fluctuations of the metallicity estimates, where the
respective measurement error associates the RGB star with an isochrone
not necessarily corresponding to its true metallicity. In these cases,
the isochrone fit can only yield a lower limit for the ``true''
stellar age. Hence, we were able to derive ages for 33 targets and
lower limits for 18 stars. For one star, T\_21, no reliable age could
be established, since its red color places it far outside the 
applicable isochrone grid. 
In 14 cases, the location of the targets above the theoretical tip RGB
coupled with their CaT metallicity makes them likely bright AGB star
candidates. As Suntzeff et al. (1993) argue, their spectroscopic metallicities 
are not expected to be underestimated by more than 0.05\,dex so 
that we did not account for their presence in our metallicity determinations. 
 Seven of these stars could also be re-identified in the
infrared point source catalog of 2MASS (Cutri et al.\ 2003). 
All of the 2MASS stars have metallicities consistent with the peak value of 
our MDF. Three of these stars are apparently younger than 5\,Gyr. 

The uncertainty of each stellar age was then obtained by means of Monte
Carlo simulations, in which  each target was varied around its uncertainty 
in metallicity and its  photometric error. 
The latter amounts to 0.02\,mag on average for both $g$ and $i$, and we also 
accounted for the r.m.s. scatter 
about the filter transformation (see eqs. 2,3), potential uncertainties in 
the reddening (0.02\,mag) and distance.    
Using these new parameters, the target was re-fit by the appropriate isochrones.  
The resulting variation of the thus derived new ages around the original value 
was then taken as a representative number of our random measurement uncertainty. 
If we were to assume an accurately determined distance and perfectly measured metallicities, 
our photometry would allow us to infer ages with about 40\% accuracy on average. 
It turns out that our metallicity uncertainties and the variation in the distance modulus 
($\pm$0.13\,mag; Bellazzini et al. 2005) contribute roughly in equal parts to the error 
budget and thus allow us to obtain individual ages with final typical random errors ranging from 25\% to 100\% 
(corresponding to 0.2--0.4\,dex in log\,$age$).

Another potential source of uncertainty in the age derivation is the a priori unknown 
[$\alpha$/Fe]-ratio in the target stars. Since dSphs are known to have 
experienced different chemical evolution histories than the Galaxy (e.g., Matteucci 2003), 
the scaled-Solar abundance ratios employed in our isochrone grids need 
not necessarily to apply to our stars (see discussions in Tolstoy et al. 2003; 
Cole et al. 2005). 
We approach this effect by representing such deviations from the standard abundance 
mixture  via the empirical relation between the overall heavy element abundance 
[$M$/H] to be used for the isochrones and the $\alpha$-element abundance from Salaris et al. (1993): 
[$M$/H]\,$\approx $\,[Fe/H]\,+\,log\,(0.638\,10$^{[\alpha/Fe]}$\,+\,0.362). 
Recent high-resolution spectroscopic data indicate that the [$\alpha$/Fe]-ratio 
in dSphs is enhanced by $\sim$0.1\,dex (Shetrone et al. 2001, 2003; Venn et al. 2004; 
Koch 2006). These data can be approximated by a linear relation of the kind 
[$\alpha$/Fe]\,$\approx$\,$-0.08-0.11$\,[Fe/H]  
in the metallicity range covered by our Leo\,II targets. 
This slight $\alpha$-enhancement leads to only a small shift towards younger ages 
w.r.t. those obtained with the scaled-Solar isochrones. All in all, the neglect of any 
$\alpha$-enhancement leads to a systematic uncertainty of $\sim$4\% in our derived ages. 

Finally, we note that the evolutionary status of a star cannot be 
unambiguously known from the CMD so that we have to identify all our targets 
with first-ascent RGB stars, unless this is excluded by the isochrone grid (as shown 
by the different symbols in Fig.~11). 
However, Bellazzini et al.\ (2005) report the detection of a secondary
bump in the RGB luminosity function at $M_V\simeq-$\,0.5\,mag ($i\simeq$\,20.4\,mag), 
which may be attributed to the AGB clump (Gallart 1998; Lee et al.\
2003), from which it is indistinguishable in practice. 
Although this feature occurs at fainter magnitudes than our target stars 
(cf. Fig.~1),  its presence indicates that our RGB
sample may contain a non-negligible admixture of stars belonging to the
AGB. 
If we simply adopt an RGB isochrone for deriving an age for a star, which was in 
fact on the AGB, we find from our isochrone grids that its age will be underestimated 
by $\sim$18\% on average (see also Cole et al. 2005).

Our final ages and their estimated uncertainties are listed in Table~3.
 
\subsection{Age-metallicity relation}

Fig.~11 shows a plot of the age-metallicity relation (AMR) obtained
from our CaT metallicities and successive isochrone fits. The first
thing to note is that our data cover a wide range in ages, ranging
from 2 to 15\,Gyr.  The AMR is essentially flat for ages above $\sim
5$\,Gyr. For the majority of our stars 
we find ages between 5 and
11.5\,Gyr with an indication of a peak around (9\,$\pm$\,2)\,Gyr.  A small number of stars is 
older than this limit (note, however, that the limits of
the applied isochrone grids extend beyond the ages of the oldest
globular clusters and of the current WMAP estimate of the age of the
Universe). Given the small sample and the uncertainties of our data, the 
inferred distribution of the ages of the stellar populations of Leo\,II is in good agreement with the 
SF histories derived from deep
CMDs (Mighell \& Rich 1996) and underscore this galaxy's extended SF history 
already from early epochs on.

The majority of the most metal-rich stars above $\sim
-1.5$~dex is identified with a number of objects younger than 5\,Gyr.
Such a seemingly sudden jump in the AMR caused by this populous group
of young stars had not been detected in Leo\,II before (see, e.g.,
Bosler et al.\ 2004). Provided the ages of this component are real and not
an artifact caused by an overestimate of their metallicity, by
Galactic interlopers, or by problems with the isochrones and the
photometry, this points to the presence of a younger population
in Leo\,II.  As summarized in the Introduction, earlier photometric
studies suggested that the main star formation occurred longer ago,
peaking at 8 or 9 Gyr (e.g., Mighell \& Rich 1998; Dolphin 2002).  Our
age estimates for the metal-rich population are consistent with the
youngest photometric ages proposed by Dolphin (2002).  If star
formation in Leo\,II did indeed last until about 2 Gyr ago, then this
dSph resembles other distant dSph companions of the Milky Way, in
particular Leo\,I, the most remote Galactic dSph satellite  (Gallart
et al.\ 1999; Held et al.\ 2000; Bosler et al.\  2004).  Furthermore,
we note that stars more metal-poor than $\sim -2$\,dex are practically
found across the full age range from 5 to 15\,Gyr, although the age
estimates for these objects provide mostly a lower boundary.  

It is clear that age derivations in Leo\,II have to be taken with
care. 
Taking our AMR at face value, our ages indicate that
Leo\,II experienced star formation continuously until about 2 Gyr ago.
It would also appear that Leo\,II did
not undergo significant enrichment during the first seven or so Gyr
after it was rapidly enriched to reach $-1.5$\,dex immediately after 
the Big Bang. 
If this was indeed the case, it would indicate
that either freshly synthesized metals must have been lost (e.g.,
through galactic winds) or that Leo\,II was constantly rejuvenated by
infall of low-metallicity gas.In this vein, a constant mean metallicity 
can be retained if metal-poor gas flows in at a same rate as the SF rate 
(e.g., Larson 1972; Lynden-Bell 1975). 
It is nevertheless unclear whether Leo\,II started 
at zero metallicity before experiencing any enrichment, or 
if the galaxy formed its stars from pre-enriched external material. 
From 7 to approximately 4 Gyr ago, the metallicity may have been slightly lower
than the mean metallicity in this galaxy. These stars then seem to have formed from more
metal-poor material than the bulk of Leo\,II's population. 
 If the
more metal-rich stars are indeed younger than 5 Gyr, then this 
appears to have been a period in which either metals were retained more
efficiently and used in subsequent star formation, or in which gas
accretion did not play a major role.   The lack of extremely
metal-poor stars, which has also been observed in all the other dSphs 
with available CaT spectroscopy so far, is very noticeable in Leo\,II as well.   More
data are needed to define the AMR more completely. 
  
\subsection{Radial or spatial trends}

Many of the Local Group dSphs exhibit radial population gradients
(e.g., Harbeck et al.\ 2001), which broadly resemble each other in the
sense that the more metal-poor (and/or older) stellar populations are
more extended compared to the more metal-rich and younger populations,
which tend to be more centrally concentrated.  This general trend has
also been detected in Leo\,II, where Bellazzini et al.\ (2005) have
shown that the predominant red clump stars are in fact
significantly more concentrated than the old blue HB and RR\,Lyrae
stars.  
The numerous red clump stars can be associated with the dominant intermediate-age
population of an age of $\sim$9\,Gyr as opposed to the older
($>$10\,Gyr) HB population (Mighell \& Rich 1996).
Hence, Bellazzini et al.\ (2005) concluded that the radial population gradient is
primarily a reflection of an age gradient. 

In Fig.~12 (top panel) we plot our CaT based metallicities versus
elliptical radius, adopting an ellipticity of 0.13 (Irwin \&
Hatzidimitriou 1995). In this representation, there is only a weak
gradient recognizable.  An error-weighted least-squares fit yields a
slope of (0.042$\pm$0.010)\,dex\,arcmin$^{-1}$ (1.5$\pm$0.4)\,dex\,kpc$^{-1}$), 
corresponding to a
change of $\sim$0.2\,dex across the face of the galaxy, which is of
the order of the measurement uncertainties.  The cumulative number
distributions in Fig.~12 do not exhibit a distinct spatial separation
of the metal-rich and metal-poor populations when subdividing  the
data at the median value of $-1.65\,$dex. 
 Although there is a weak indication
of an excess of metal-rich stars at about 4--5$\arcmin$, we cannot
exclude the possibility that both populations have the same spatial
distribution.  A K-S test shows a probability of 76\% (1.2$\,\sigma$)
for this possibility.  This result underscores the suggestion by
Bellazzini et al.\ (2005) that, despite the presence of metallicity
variations, the main driver of a population gradient in Leo\,II seems
to be the age.  

On the other hand, in terms of an age gradient, stars younger and
older than 10\,Gyr exhibit the same spatial distribution with a K-S
probability of 98\%. Also when confining the data to only those stars
of the age peak around 9\,Gyr, the probability of a same radial
distribution is comparably high, but may be hampered by the sparse
sampling of older stars.  In fact we cannot exclude the hypothesis
that the populations of different ages are equally distributed 
at the 2$\,\sigma$-level.
Considering the uncertainties associated with RGB age dating and with
the small number statistics of our study, a better way of
investigating the presence of any age gradients is by using stars that
are secure age tracers such as HB stars, red clump stars, and stars along the
vertical red clump (see, e.g., Grebel 1997; Stetson et al.\ 1998; Harbeck et
al.\ 2001).

\section{Summary and Discussion} 

After Leo\,I, Leo\,II is the most remote dSph companion of the Milky Way. 
Its prevailing intermediate-age stellar population as
well as its significant population of old stars make it an ideal testbench
to study  star formation that extended over many billions of years.
Existing photometric and spectroscopic data as well as our own results
suggest that SF activity may have ceased only 2 Gyr ago (Dolphin
2002; Bosler et al.\ 2004).  In the framework of a VLT Large Programme
tailored to investigate kinematical and chemical evolutionary aspects
of Galactic dSphs we have determined CaT metallicities for 52 red
giants in this system, thereby doubling the largest previously
published data sets (Bosler et al.\ 2004; B06). Our targets cover the
entire area of the galaxy, thus allowing us to efficiently trace
potential spatial metallicity variations.  

By employing the well-established CaT technique we showed that
Leo\,II's MDF peaks at $-$1.74\,dex (CG scale), in agreement with
previous spectroscopic and photometric estimates (e.g., Bellazzini et
al.\ 2005 and references therein). We infer 
a range of metallicities of $\sim 1.3$ dex.  We emphasize that
Leo\,II does not contain any extremely metal-poor stars of comparably
low metallicities as found in the Galactic halo field stars. 
The
lowest metallicities inferred in our study are similar to those of
metal-poor Galactic globular clusters.  
In the compilation  of Venn et al. (2004), 7\% of the (kinematically selected) halo stars 
lie in the regime of [Fe/H]$\le-2.5$ and only 1\% lie below $-3$\,dex. 
If these numbers were to be considered also  
representative of the dSphs' MDFs, we would expect to find about three to four stars 
more metal poor than $-2.5$\,dex in Leo\,II. 
A similar lack of extremely
metal-poor stars has also been observed in other Galactic dSphs (e.g.,
Shetrone et al.\ 2001; Pont et al.\ 2004; Bosler et al.\ 2004; Tolstoy
et al.\ 2004; Koch et al.\ 2006a,b).  While these studies all tried to cover the full
color range of the RGB to ensure that stars of very low or very high
metallicities, if present, would not be missed, 
the aforementioned deficiency might still be blamed on
small number statistics.  

For want of adequately sophisticated chemical
evolution models of the Leo\,II dSph  itself, we compared the observed
MDF to the model predictions for other Galactic dSphs from LM04.
These models include the effects of supernovae of type Ia and II and
galactic winds.  None of these models really fits Leo\,II's MDF well,
which may not be surprising given that each Galactic dSph has
experienced its own unique evolutionary history (e.g., Grebel 1997,
1999).  LM04's model for the Carina dSph approximates the MDF of
Leo\,II best, and although Car experienced strongly episodic star
formation its age range is comparable to that of Leo\,II.  

The range in [Fe/H] covered by our MDF, combined with the fact that
there is no obvious trend of metallicity with the color of the red
giants indicate that there is a broad range of ages present in
Leo\,II, a fact already known from earlier photometric CMD analyses.
By means of isochrone fitting using our spectroscopic metallicities as
an additional constraint, we could show that the ages span the full
range  from 2 to 15\,Gyr. There is a considerable fraction of
``younger'' and more metal-rich stars with ages of a few Gyr
in our data, but one cannot exclude the possibility that these may in
fact be AGB contaminants.  

Our AMR suggests constant star formation across our entire age range, although 
we note that our measurement accuracy and the age resolution would not 
allow us to detect any potential gaps in Leo\,II's SF history.  
The metallicity remained fairly constant for most of the active
periods of Leo\,II, indicating early star formation from pre-enriched
gas and possibly continued rejuvenation by little-enriched gas until
some 5 Gyr ago.  However, we caution that this qualitative
interpretation is subject to the many uncertainties affecting the
derivation of ages. 

There appears to be no radial population gradient from the
spectroscopic point of view, at least not in metallicity and at best
weakly indicated in our inferred age distribution.  Bellazzini et
al.\ (2005) show that Leo\,II follows the common trend of containing
more centrally concentrated intermediate-age populations as compared
to the old component.  
 In the case of Leo\,II the main driver
of the population gradient appears to be age, since we have
demonstrated that the AMR shows little evidence for metallicity
variations (i.e., enrichment) across most of the age range.

\acknowledgments

A.K.\ and E.K.G.\ are grateful for support by the Swiss National Science
Foundation (grant 200020-105260).  M.I.W.\ acknowledges the Particle
Physics and Astronomy Research Council for financial support.  
We would like to thank Gustavo Lanfranchi for sending us his models 
and the anonymous referee for a very helpful report.  

Funding for the SDSS and SDSS-II has been provided by the Alfred P.\
Sloan Foundation, the Participating Institutions, the National Science
Foundation, the U.S.\ Department of Energy, the National Aeronautics
and Space Administration, the Japanese Monbukagakusho, the Max Planck
Society, and the Higher Education Funding Council for England.  The
SDSS Web Site is http://www.sdss.org/.

The SDSS is managed by the Astrophysical Research Consortium for the
Participating Institutions.  The Participating Institutions are the
American Museum of Natural History, Astrophysical Institute Potsdam,
University of Basel, Cambridge University, Case Western Reserve
University, University of Chicago, Drexel University, Fermilab, the
Institute for Advanced Study, the Japan Participation Group, Johns
Hopkins University, the Joint Institute for Nuclear Astrophysics, the
Kavli Institute for Particle Astrophysics and Cosmology, the Korean
Scientist Group, the Chinese Academy of Sciences (LAMOST), Los Alamos
National Laboratory, the Max-Planck-Institute for Astronomy (MPIA),
the Max-Planck-Institute for Astrophysics (MPA), New Mexico State
University, Ohio State University, University of Pittsburgh,
University of Portsmouth, Princeton University, the United States
Naval Observatory, and the University of Washington.

This research has made use of NASA's Astrophysics Data System
Bibliographic Services.

\clearpage
\begin{figure}
\includegraphics[width=14cm]{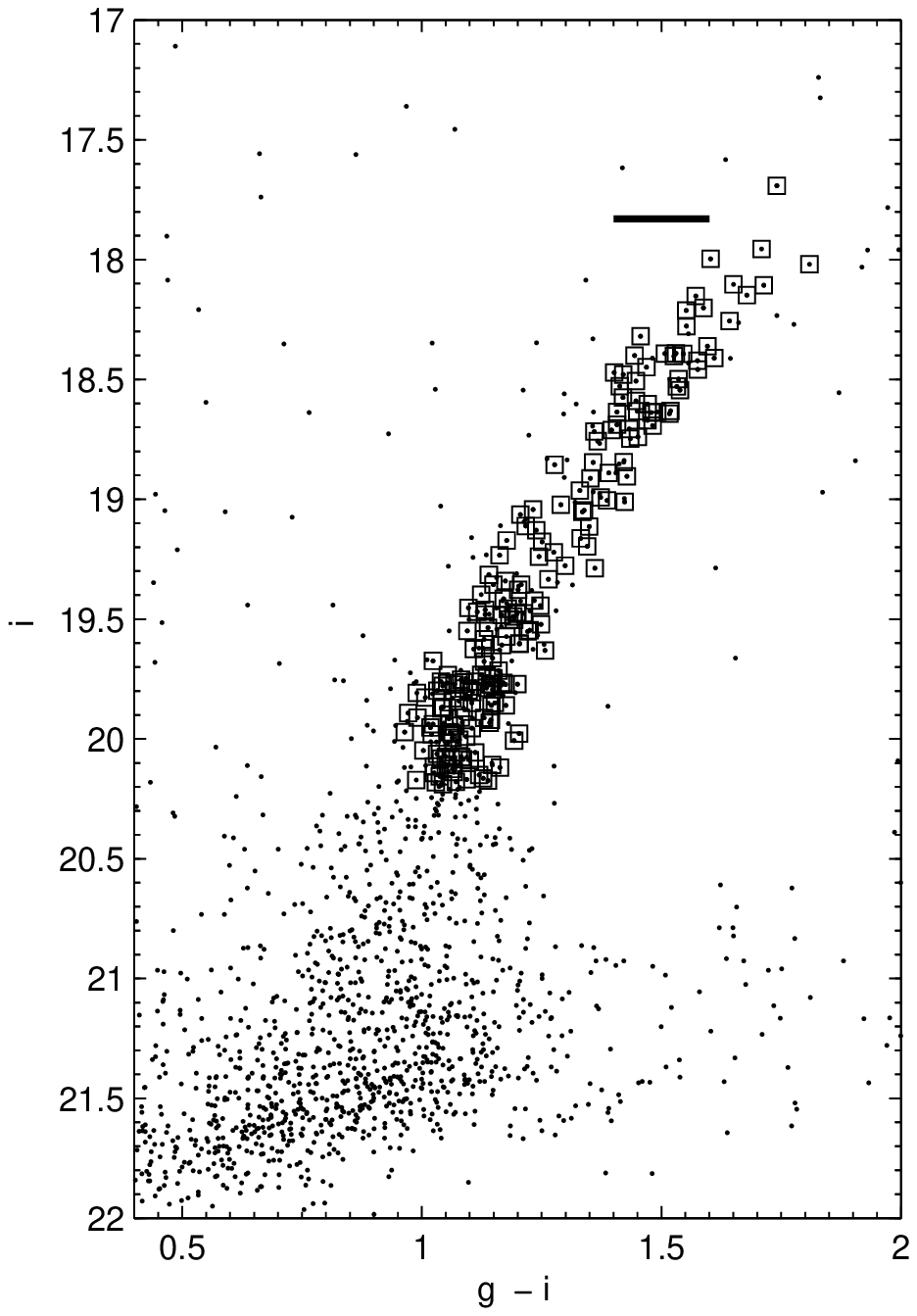}
\caption{The color-magnitude diagram of the upper red giant branch of 
Leo\,II.  Stars for which we obtained spectra are marked by open
squares.  The $g$ and $i$ photometry shown is in the SDSS photometric 
system and was transformed into this system from INT/CASU $g'$ and $i'$
magnitudes. Also indicated by the black solid line is the location of the TRGB 
according to Bellazzini et al. (2005). 
}
\end{figure}

\begin{figure}
\plotone{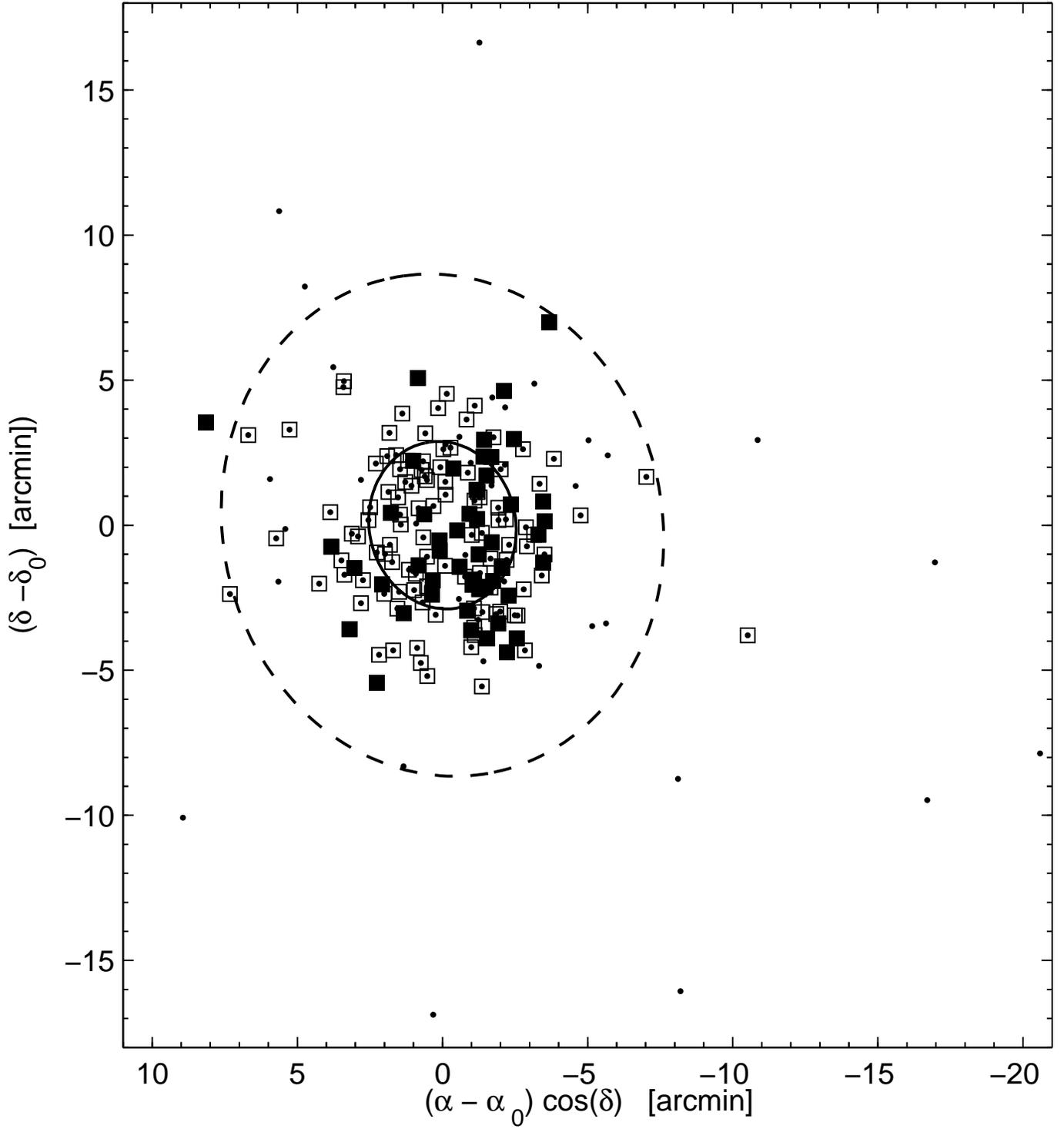}
\caption{Location of our 197 targets (shown as small dots) in a right
ascension versus declination plot after subtracting the central
coordinates $\alpha_0$ and $\delta_0$ of Leo\,II.  Open squares mark  radial velocity 
members.  Stars indicated by filled squares have sufficient
signal to noise to permit us to derive their metallicities.  
The dashed line shows Leo\,II's nominal tidal radius (8$\farcm$9.)}
\end{figure}

\begin{figure}
\plotone{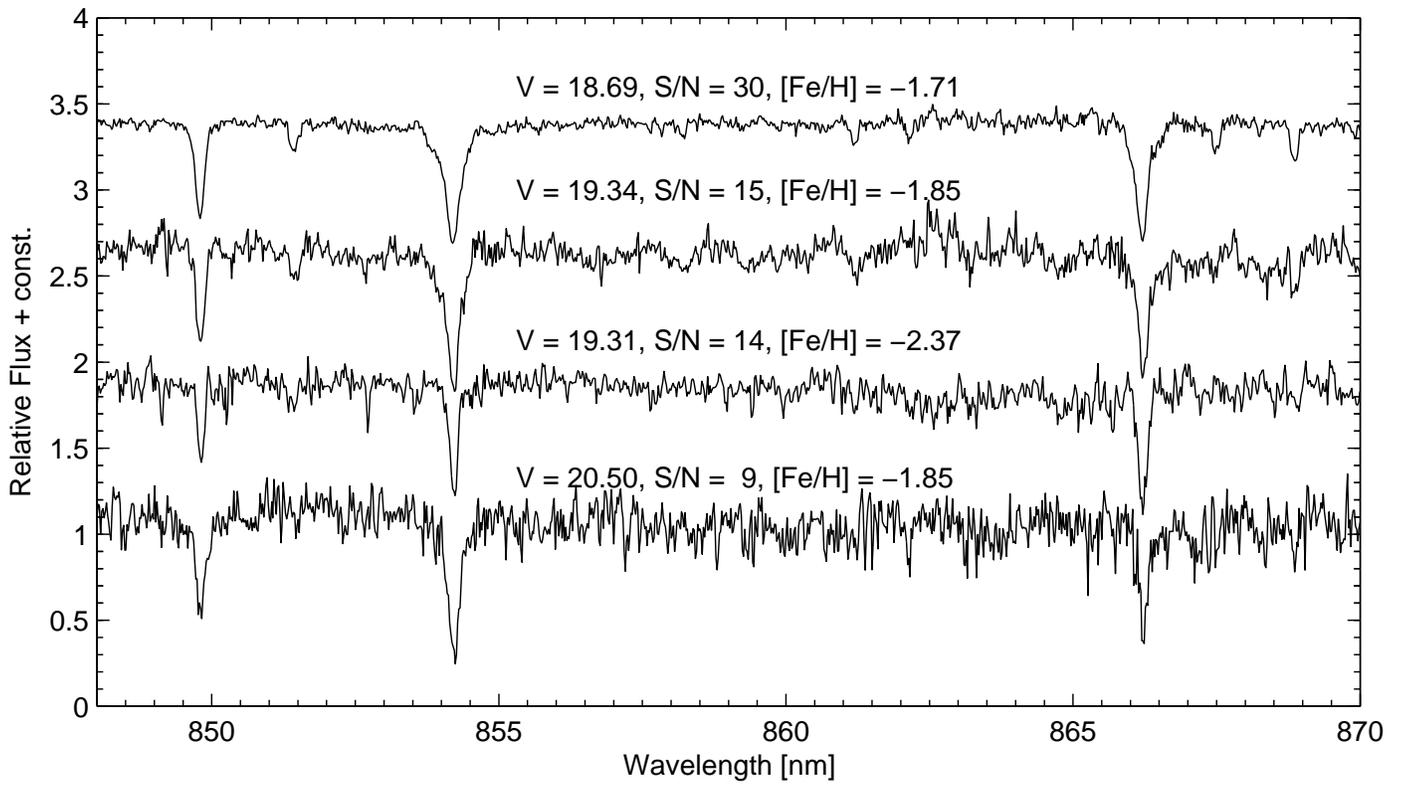}
\caption{Sample spectra showing some of the more metal-poor stars 
covering a range of different magnitudes and signal-to-noise ratios.}
\end{figure}

\begin{figure}
\plotone{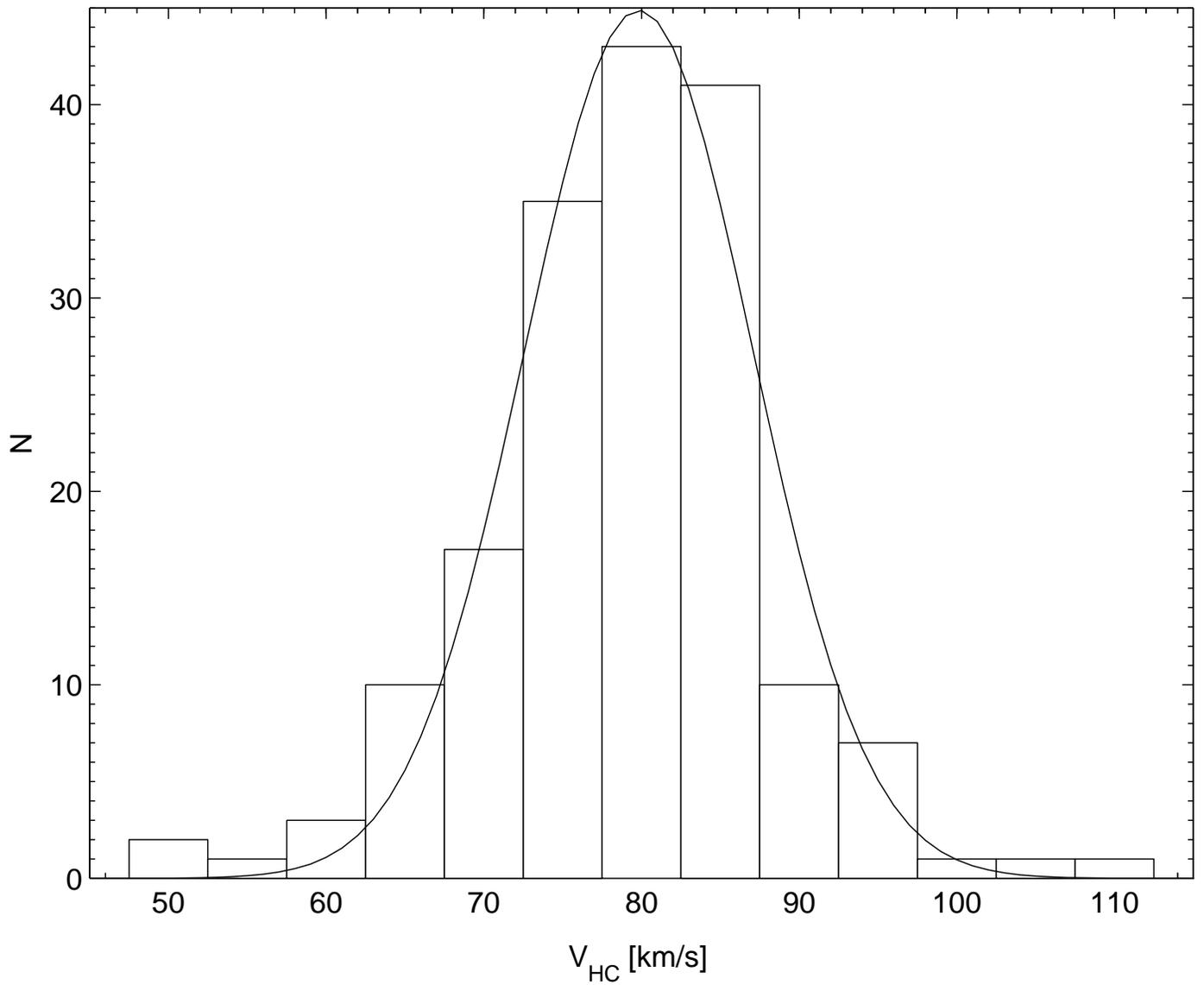}
\caption{Radial velocity histogram of our target stars, most of which
are red giants in Leo\,II. The solid line is the best-fit 
Gaussian to this distribution. }
\end{figure}

\begin{figure}
\plotone{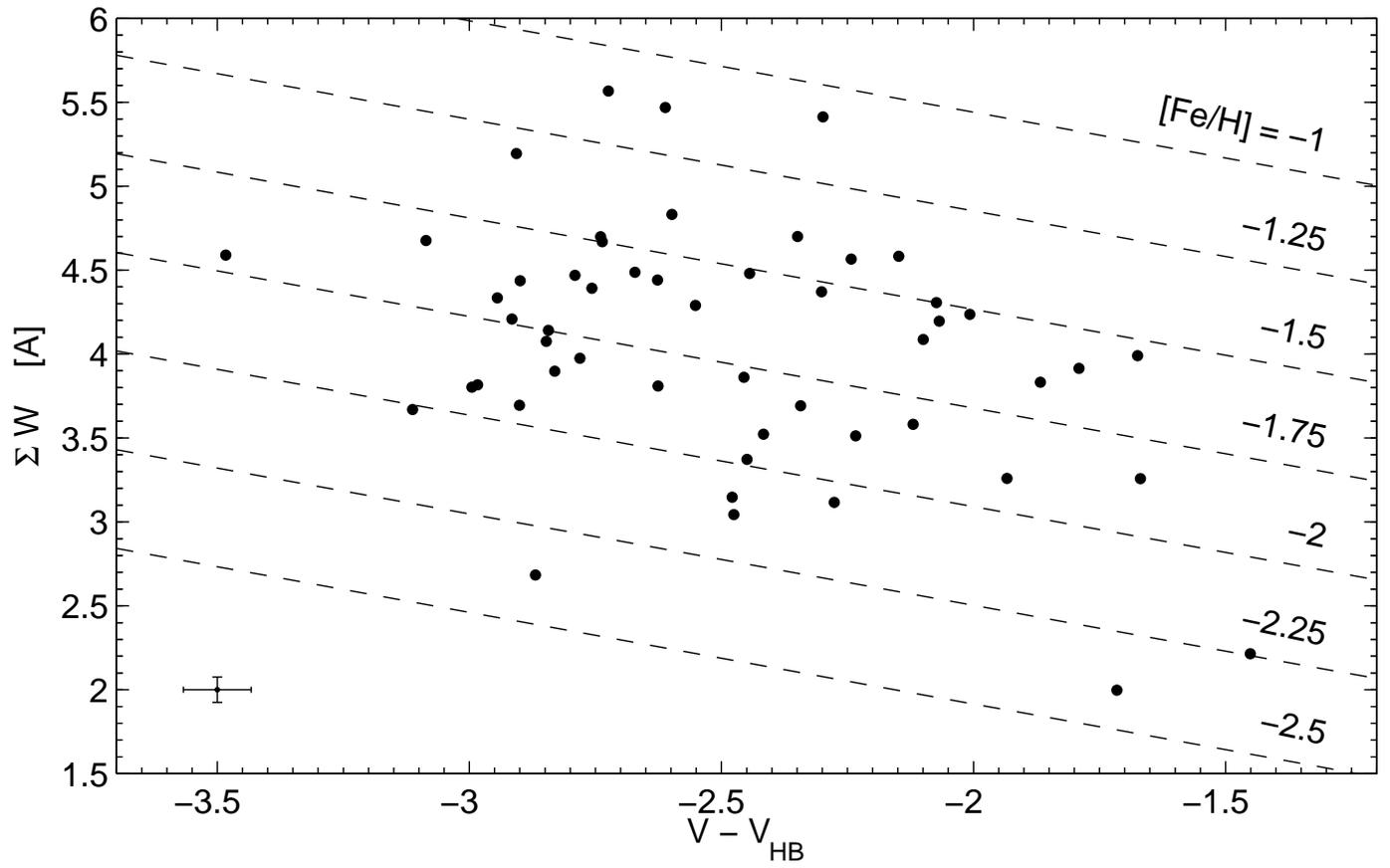}
\caption{Distribution of our targets in the ($W$, $V-V_{HB}$)-plane. 
Also ahown are lines of constant metallicity. A typical errorbar is indicated 
in the lower left.}
\end{figure}

\begin{figure}
\plotone{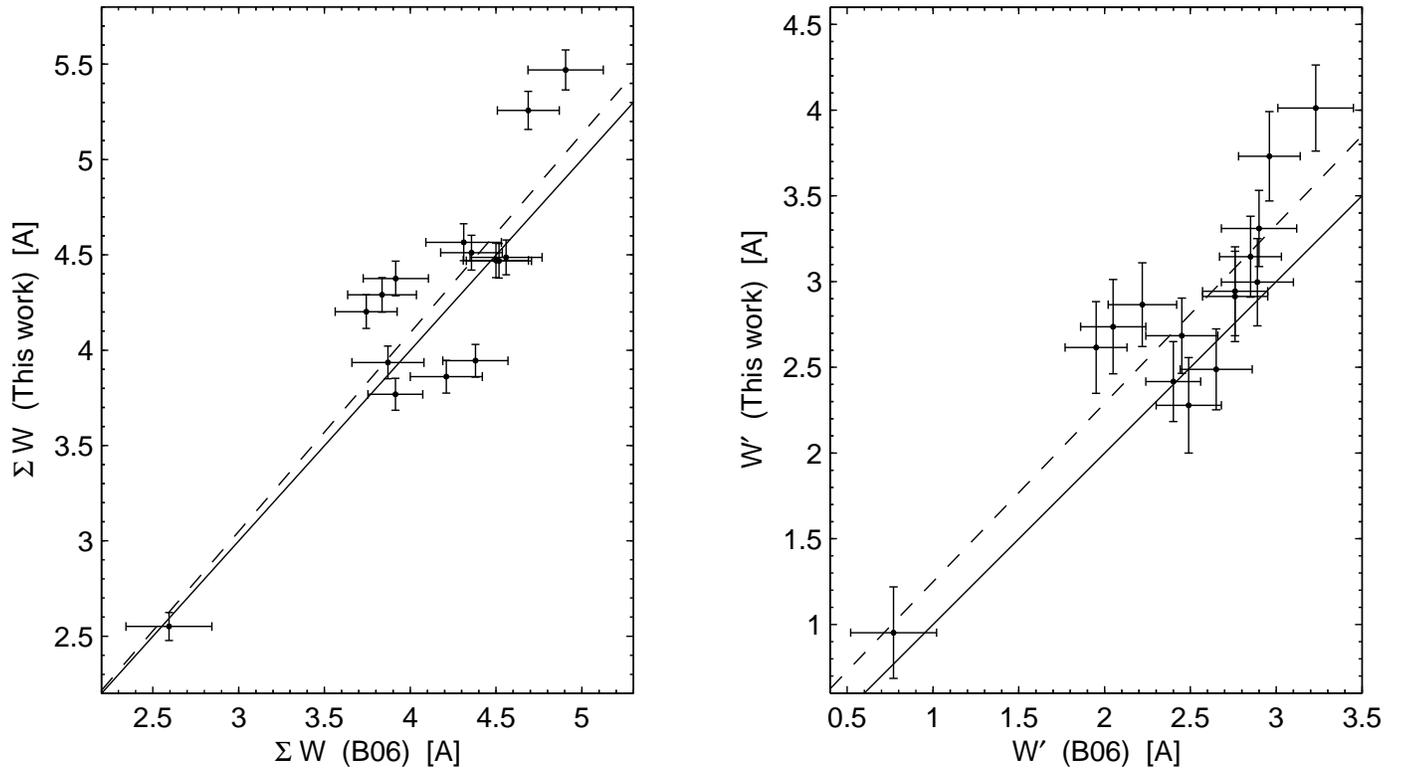}
\caption{The CaT linestrength (left panel) and reduced width (right) as measured in this work versus 
the data from B06.  The solid and dashed lines are unity and the 
best-fit relations, respectively.
Errorbars indicate formal 1$\sigma$ uncertainties, also accounting 
for the uncertainties in the calibrations (eqs. 5, 6).}
\end{figure}

\begin{figure}
\plotone{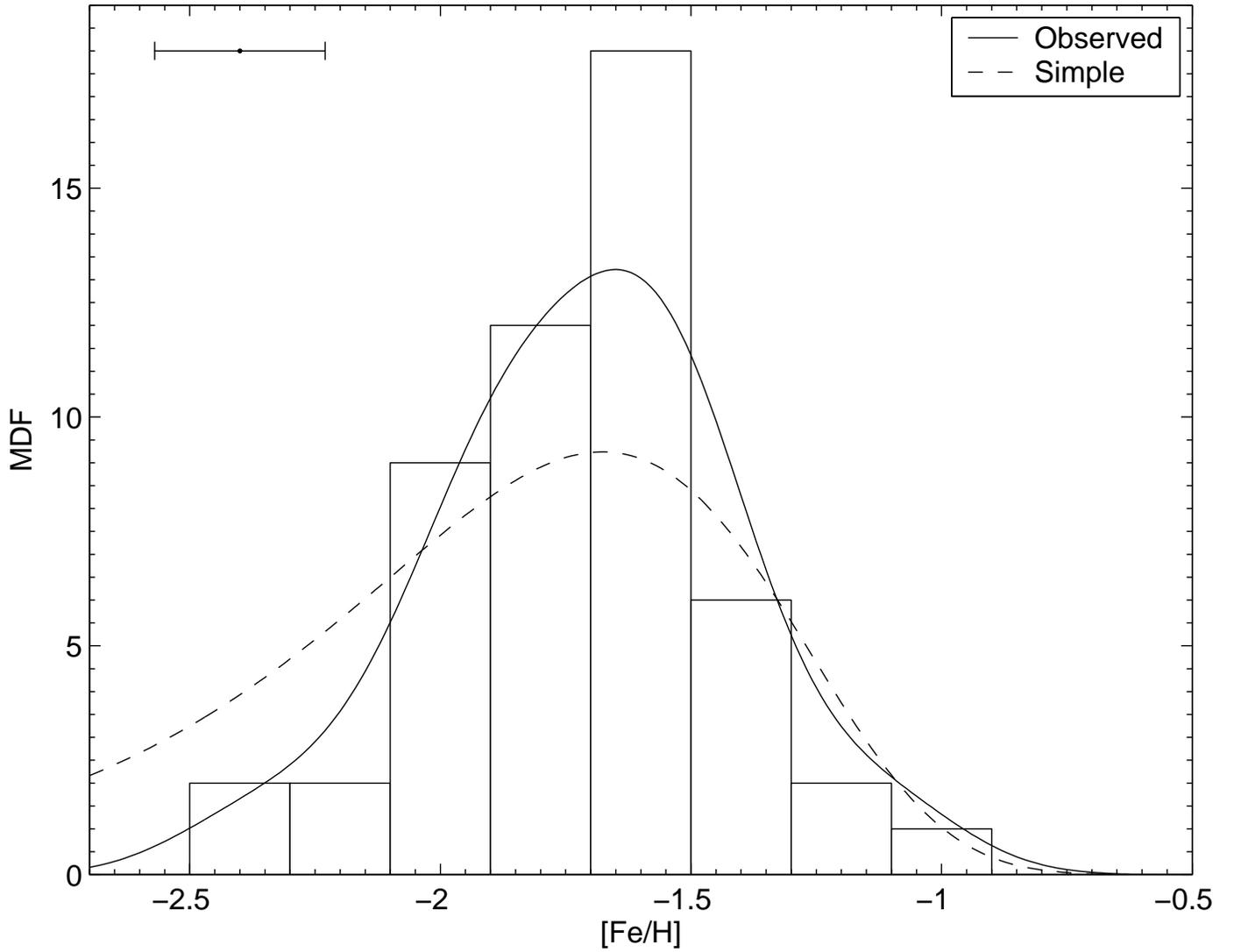}
\caption{Metallicity distribution function on the scale of CG for the 
52 Leo\,II member stars for which we were able to measure
metallicities. Overplotted as a solid line is the MDF convolved with  
observational uncertainties, also indicated in the top left corner by 
a representative 1$\sigma$ errorbar. 
Additionally shown is a modified simple model (``Simple'') of chemical evolution, 
which strictly holds only for long-lived stars. The model was scaled to the 
same number of stars.}
\end{figure}

\begin{figure}
\plotone{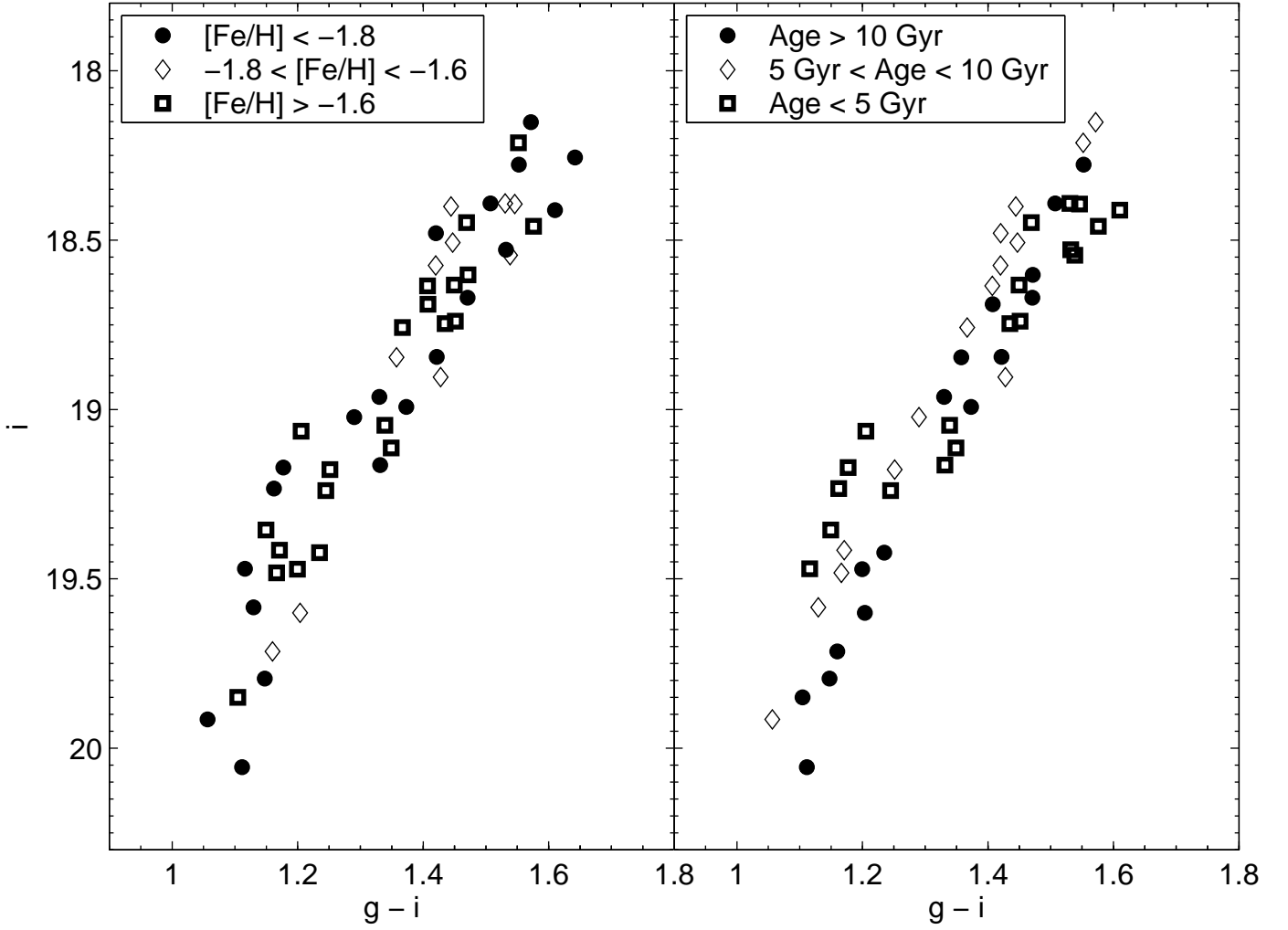}
\caption{Color-magnitude diagrams showing the targeted red giants in
Leo\,II.  In the left panel, the different symbols indicate three
different metallicity ranges.  In the right panel, the symbols show 
three different age ranges.}
\end{figure}

\begin{figure}
\plotone{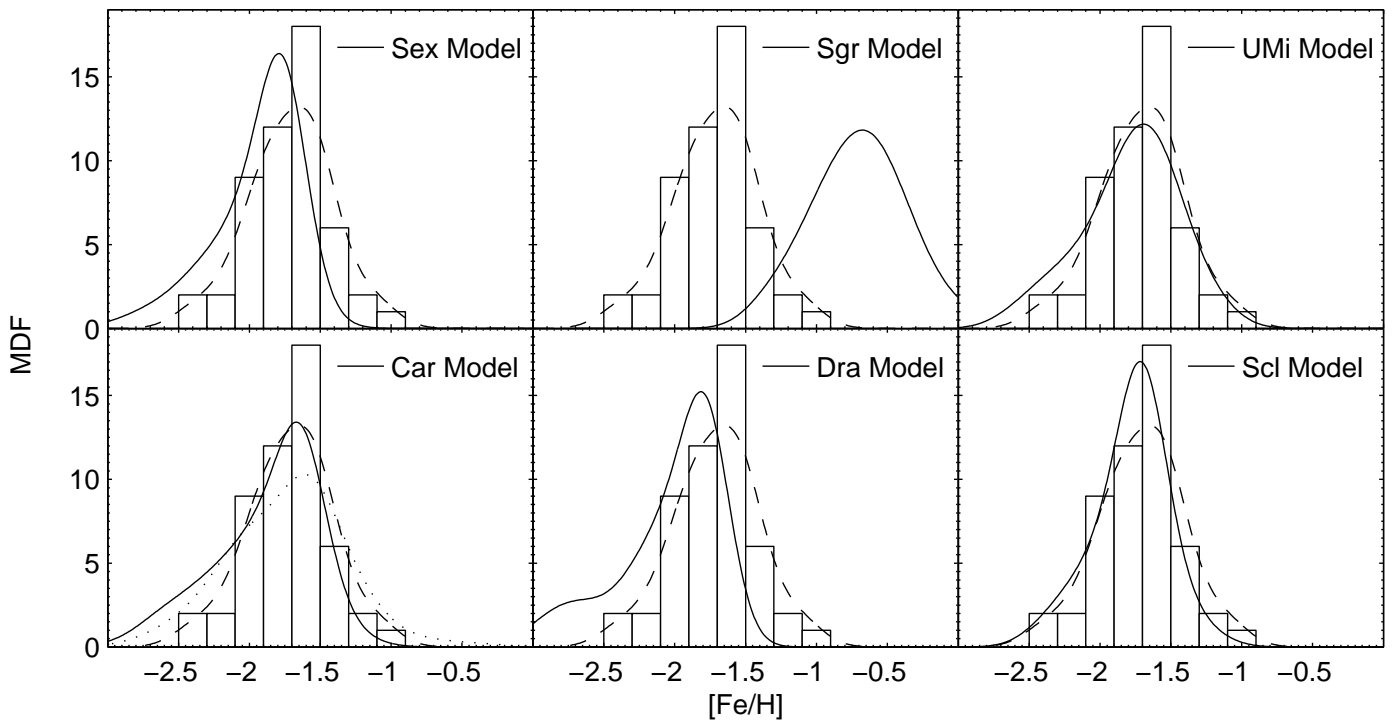}
\caption{Observed MDF versus different models: The histogram and 
dashed curve show the observed MDF of Leo\,II (convolved with 
measurement errors).  The solid lines in each panel are the 
predictions for six dSphs from the models of LM04, also convolved 
with the observational errors from the present study.  
The lower left panel also superimposes the observed 
MDF of Carina from Paper\,I (dotted line). 
All distributions were normalized to the same number of stars.}
\end{figure}
\begin{figure}
\plotone{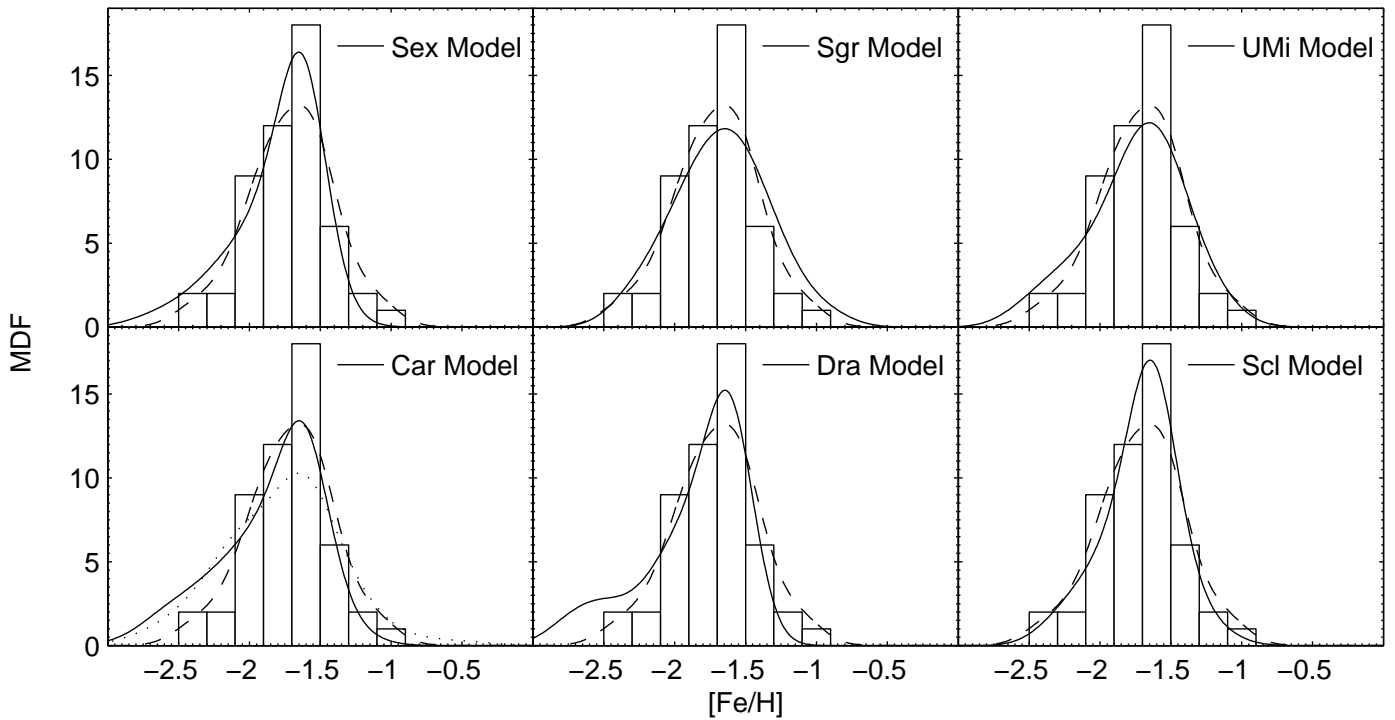}
\caption{Same as Fig.~7, but all models were shifted such in
metallicity that the location of their peaks coincide with the peak of 
the observed MDF of Leo\,II.  }
\end{figure}

\begin{figure}
\plotone{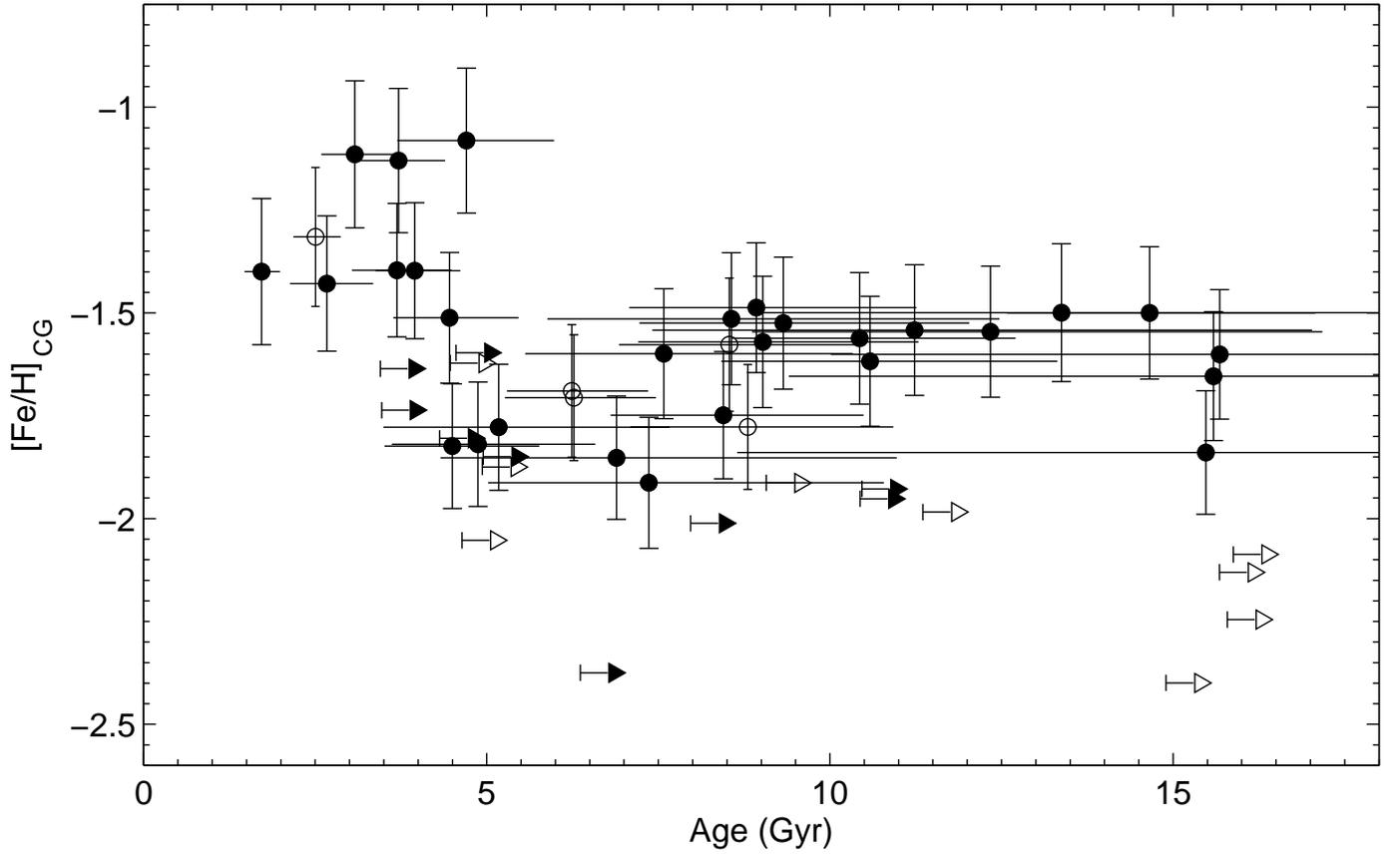}
\caption{The age-metallicity relation inferred from our CaT data and
isochrone fits to the SDSS photometry.  The arrows indicate lower limits 
where no reliable age could be determined. The open symbols denote 
targets that lie above the theoretical tip of the RGB of their 
respective metallicity. See text for details.}
\end{figure}

\begin{figure}
\plotone{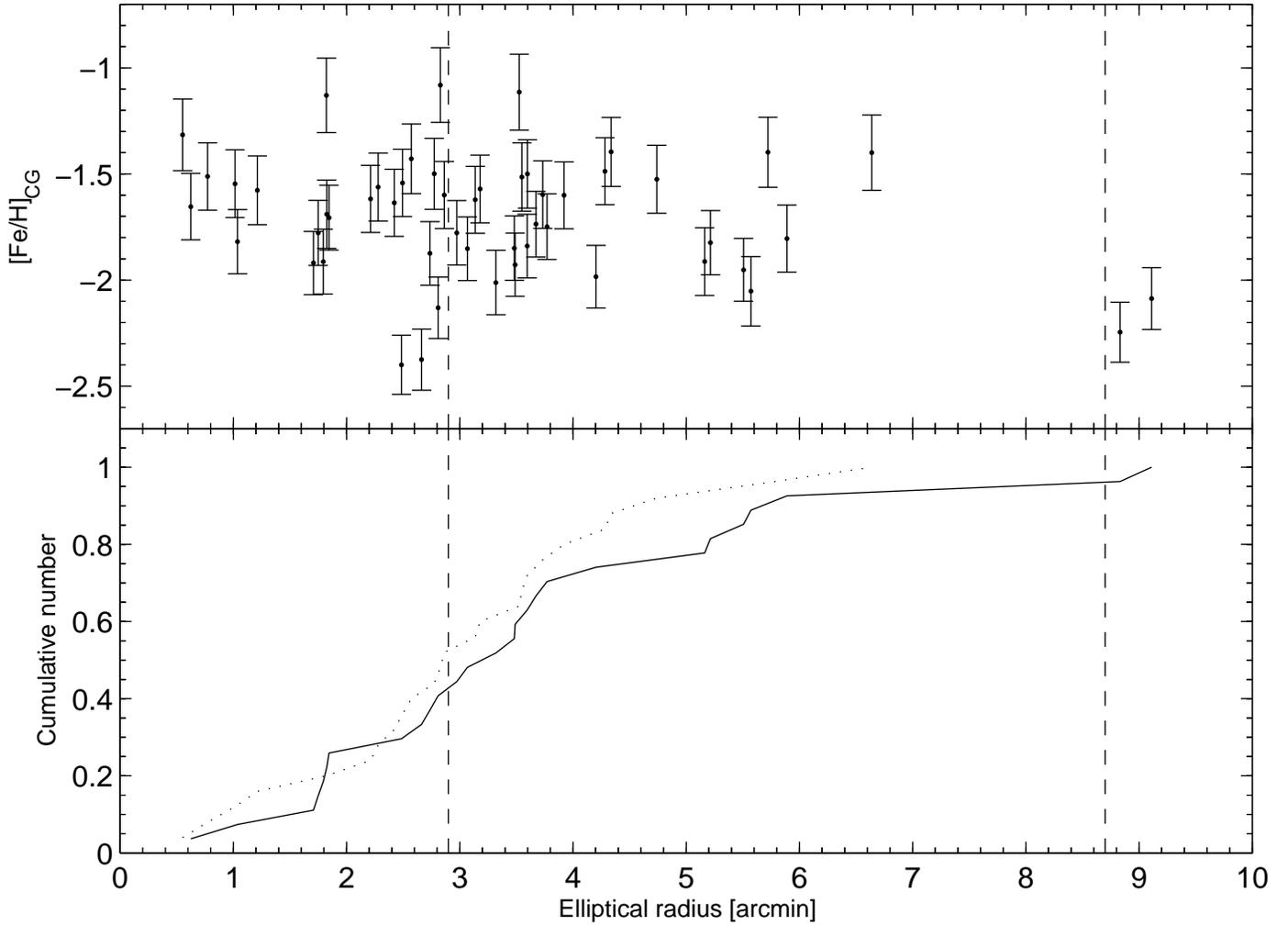}
\caption{Top panel: Metallicity versus distance from the center of
the galaxy. The bottom panel shows the cumulative number fraction, 
separately for the metal-poor ([Fe/H]$<-1.65$, solid line) and 
metal-rich ([Fe/H]$\ge-1.65$, dotted line) targets.  
The dashed vertical lines represent the location of Leo\,II's core  
and nominal tidal radius, respectively. 
There is only a weak indication of a radial metallicity gradient.}
\end{figure}

\clearpage

\begin{table}
\begin{center}
\caption{Observation log}
\begin{footnotesize}
\begin{tabular}{lcc}
\hline
\hline
     &  & Total Exposure Time \\ 
Date & Field (Configuration) & [s] \\
\hline
2003 Feb 21 & Center    &  1436 \\ 
2003 Feb 22 & Center    & 18000 \\
2003 Mar 04 & Center\_2	& 18000 \\
2003 Mar 05 & Offset    & 14400 \\
2004 Feb 21 & 2 (a)        & 10800 \\
2004 Feb 22 & 2 (a)        &  3600 \\
2004 Feb 23 & 2 (a)        & 10800 \\
2004 Feb 24 & 2 (b)        & 10020 \\
2004 Feb 26 & 4 (a)        & 18300 \\
2004 Feb 27 & 4 (b)        & 15166 \\
2004 Feb 28 & 4 (b)        & 14700 \\
\hline
\end{tabular}
\end{footnotesize}
\end{center}
\end{table}

\begin{table}
\begin{center}
\caption{Observed fields in Leo\,II}
\begin{footnotesize}
\begin{tabular}{clcc}
\hline
\hline
Field & $\alpha$ (J2000.0) & $\delta$ (J2000.0) \\
\hline
Center	  & 11 13 23.9  & 22 14 33 \\
Center\_2 & 11 13 32.8  & 22 06 30 \\
Offset	  & 11 12 42.6  & 22 02 21 \\
2	  & 11 13 36.2  & 22 11 08 \\
4	  & 11 12 45.9  & 21 58 35  \\ 
\hline
\end{tabular}
\end{footnotesize}
\end{center}
\end{table}
\begin{table}
\begin{center}
\caption{Measured properties of radial velocity members in Leo\,II}
\begin{footnotesize}
\begin{tabular}{lcccccccccc}
\hline
\hline
Star & $\alpha$ (J2000) & $\delta$ (J2000) & r\,[$\arcmin$] 
 & V & g & g$-r$ & g$-i$ & W'\,[\AA]  & [Fe/H]$_{\mathrm{CG}}$ & Age [Gyr] \\
\hline
T\_11 & 11 13 12.2 & 22 16 09 &  8.83 & 20.72 & 21.17 &  0.75 &  1.11 &  1.42 $\pm$  0.18 & $-$2.25 $\pm$  0.12  & 15.8$_{-5}^{+6}$ \\ 
T\_13 & 11 13 35.8 & 22 09 35 &  1.84 & 18.69 & 19.43 &  1.22 &  1.74 &  2.69 $\pm$  0.32 & $-$1.71 $\pm$  0.18  & 6.3$_{-1}^{+1}$ \\ 
T\_16 & 11 13 13.9 & 22 08 50 &  3.32 & 19.06 & 19.72 &  1.09 &  1.57 &  1.97 $\pm$  0.30 & $-$2.01 $\pm$  0.16  & 8.0$_{-1}^{+2}$ \\ 
T\_19 & 11 13 23.0 & 22 09 23 &  1.21 & 19.08 & 19.76 &  1.12 &  1.55 &  2.99 $\pm$  0.30 & $-$1.58 $\pm$  0.17  & 8.5$_{-2}^{+2}$ \\ 
T\_21 & 11 13 22.7 & 22 08 09 &  1.71 & 19.19 & 19.90 &  1.17 &  1.64 &  2.19 $\pm$  0.28 & $-$1.92 $\pm$  0.16  & \dots \\ 
\hline
\end{tabular}
\end{footnotesize}
\end{center}
{\footnotesize Note. --- This Table is published in its entirety in the electronic edition of the {\it Astronomical Journal}.
A portion is shown here for guidance regarding its form and content.
$r$ and W' denote the elliptical radius and the reduced CaT width. 
Colors are given in the photometric system of the SDSS.  
[Fe/H] and the respective uncertainty are given following eq. 7.}
\end{table}
\clearpage

\end{document}